\documentclass[twocolumn]{aastex6}

\received{June 20, 2017}
\revised{September 10, 2017}
\accepted{XXXX}

\shorttitle{hot gaseous halo around the Milky Way}
\shortauthors{Li \& Bregman}

\begin{document}
\title{The Properties of the Galactic Hot Gaseous Halo from X-ray Emission}

\author{Yunyang Li}
\affiliation{Peking University, Beijing 100871,China}
\email{liyunyang@pku.edu.cn}

\author{Joel Bregman}
\affiliation{University of Michigan, 1085 S.University Ave. Ann Arbor, 40109}



\begin{abstract}
The extended hot X-ray emitting gaseous halo of the Milky Way has an optical depth $\sim1$ for the dominant emission lines of \ion{O}{7} and \ion{O}{8}, which are used to infer the halo properties. To improve on halo gas properties, we treat optical depth effects with a Monte-Carlo radiative transfer model, which leads to slightly steeper density profiles ($\beta \approx 0.5$) than if optical depths effects were ignored. For the preferred model where the halo is rotating on cylinders at $180$ km s$^{-1}$, independent fits to both lines lead to identical results, where the core radius is $2.5$ kpc and the turbulent component of the Doppler b parameter is $100-120$ km s$^{-1}$; the turbulent pressure is $20\%$ of the thermal pressure.  The fit is improved when emission from a disk is included, with a radial scale length of $3$ kpc (assumed) and a fitted vertical scale height of approximately $1.3$ kpc. The disk component is a minor mass constituent and has low optical depth, except at low latitudes. The gaseous mass is $3-4\times10^{10}\,M_{\odot}$ within $250\,\mathrm{kpc}$, similar to our previous determinations and significantly less than the missing baryons of $1.7\times10^{11}\,M_{\odot}$.
\end{abstract}




\section{Introduction} \label{sec:intro}

Galaxy formation through spherical accretion naturally produces an accretion shock with gas near the virial temperature $T_{\mathrm{vir}}$ \citep{1991ApJ...379...52W}. This picture is modified by radiative losses and non-spherical accretion, leading to cooled gas but still with a hot halo present. Simulations 
\citep[e.g.,][]{2006ApJ...650..560C,2009MNRAS.392...77T,2009ApJ...694L.123K,2006MNRAS.366..499D,2010MNRAS.407.1403C,2012ApJ...759..137J}
show that hot halos are prominent in more massive galaxies \citep[$M_{\mathrm{halo}}\geq10^{11.4}M_{\odot}$,][]{2005MNRAS.363....2K}, where the cooling time is longer. The mass of the hot gaseous halo is modified significantly by feedback, especially in the region where $t_\mathrm{cool} < t_\mathrm{Hubble}$. The hot halo may be important from the census of metals and the census of baryons. Compared to cosmological observation \citep{2013ApJS..208...19H}, most of the baryons and metals are missing from galaxies \citep[see][for a review]{2007ARA&A..45..221B}. It is suggested that a significant fraction of the missing baryons and metals lie in an extended hot halo, out to, or beyond $R_{\mathrm{vir}}$ \citep{1998ApJ...503..518F,2005Natur.433..495N,2005ApJ...631..856W,2006ApJ...645..179W,2006ApJ...644L...1S,2006ApJ...639..590F,2007ApJ...665..247W}.

We detect extended halos around some external galaxies, both in field early-type galaxies \citep{1985ApJ...293..102F,2001MNRAS.328..461O,2010ApJ...715L...1M}, and in spiral galaxies \citep{1997ApJ...485..159B,2006A&A...448...43T,2008MNRAS.390...59L,2011ApJ...737...22A,2012ApJ...755..107D,2013ApJ...772...98B,2013ApJ...772...97B,2015MNRAS.449.3527W}. However, for early-type galaxies, their structure and content are likely to be affected by merger history and the interaction with intergroup/intracluster medium in which most large ellipticals reside \citep{2010ApJ...715L...1M}. Even for isolated early-type galaxies and late-type galaxies, their X-ray luminosities fall below the detection limit at radii beyond $\sim 0.1 R_{\mathrm{vir}}$ \citep[]{2016MNRAS.455..227A}. In the contrary, observing from the interior provides us an unparalleled opportunity to study the hot halo around the Milky Way.

At the Milky Way's virial temperature ($\approx2\times10^6\,\mathrm{K}$), \ion{O}{7} and \ion{O}{8} lines act as ideal tracers for the hot gas, because their emissivities are most sensitive at this temperature \citep{1993ApJS...88..253S}. 
By probing the \ion{O}{7} absorption against more than $\sim 30$ AGNs and X-ray binaries, the local hot medium is studied in absorption lines \citep{2002ApJ...573..157N,2004ApJ...617..232M,2005ApJ...624..751Y,2005ApJ...631..856W,2006ApJ...645..179W,2007ApJ...665..247W,2010PASJ...62..723H,2012ApJ...756L...8G,2012ApJ...746..166Y,2013ApJ...770..118M,2014ApJ...785L..24F,2016ApJ...822...21H,2016ApJ...828L..12N} and is suggested to be associated with the Galaxy \citep[i.e., rather than with the intergroup hot gas,][]{2006ApJ...644..174F,2007ApJ...669..990B}. The studies on diffuse Milky Way emission from empty fields start since the {\it ROSAT} sky survey \citep{1995ApJ...454..643S,1997ApJ...485..125S} and continue with more data from subsequent observatories \citep{2002ApJ...576..188M,2009ApJ...690..143Y,2012ApJ...756L...8G,2013ApJ...773...92H}. Combined with absorption data, the constraints on the halo property are obtained, which are accordant with external galaxy observation.

The optical depth effects of the hot gas are important in line analysis, and the studies are carried out progressively for absorption lines. 
 \cite{2013ApJ...770..118M} use \ion{O}{7} absorption towards 29 targets from the {\it XMM-Newton} Reflection Grating Spectrometer (RGS) archival data, and they include the curve-of-growth analysis to correct the column density for optical depth effects. 
Assuming that the turbulence of the halo is $b=150\,\mathrm{km\,s^{-1}}$, the authors obtain $n_0=0.46^{+0.74}_{-0.35}\,\mathrm{cm}^{-3}$, $r_c=0.35^{+0.29}_{-0.27}\,\mathrm{kpc}$, $\beta=0.71^{+0.13}_{-0.14}$ which are slightly different from the ones obtained via an optically thin analysis and lead to a smaller baryon mass estimation ($1.2\times10^{10}\,M_{\odot}$, compared to $2.4\times10^{10}\,M_{\odot}$ in the optically thin case). 
In the study of a more comprehensive set of absorption lines, \cite{2015ApJS..217...21F} fit the Doppler-$b$ parameters which are consistent with $98.0\pm19.4\,\mathrm{km\,s^{-1}}$. 
Studies probing the column density and Doppler-$b$ parameter are also carried out through the absorption line ratios approach. \cite{2012ApJ...756L...8G} measure the equivalent width ratio between \ion{O}{7} $\mathrm{K}\alpha$ and $\mathrm{K} \beta$ lines and place a constraint on the Doppler-$b$ for each sight-line. 
Their results indicate that most ($6$ out of $8$ of their sample) \ion{O}{7} $\mathrm{K} \alpha$ lines are indeed saturated with the Doppler-$b=95.0\pm17.1\mathrm{\,km\,s}^{-1}$.
Similar analysis is also performed recently by \cite{2016ApJ...828L..12N} for low and high galactic latitude sight-lines and they find the mean Doppler-b values to be $125\,\mathrm{km\,s^{-1}}$ and $95\,\mathrm{km\,s^{-1}}$, respectively. \cite{2017A&A...605A..47N} perform an elaborate analysis on the local absorption spectra of blazar PKS 2155-304 to study the ionization of the hot halo as well as lower temperature plasmas. Their results for the \ion{O}{7} lines indicate a Doppler-$b$ of about $80\,\mathrm{km\,s^{-1}}$. 

The emission properties of the halo are studied in \cite{2015ApJ...800...14M} (hereafter, MB15), in which the authors use emission map of \ion{O}{7} and \ion{O}{8} from \cite{2012ApJS..202...14H} and fit the observation to the halo model. 
They calculate the line intensities with an optically thin model, and correct the optical depth effects by a mean scattering term (i.e., a single-scattered scenario). 
By fitting the data with and without the optical depth effect separately, they reach similar but different results, from which they well constrain the halo model and claim that the optical depth effects are partially considered. 
However, for their best-fit result of \ion{O}{8}, $n_0r_c^{3\beta}=1.5\,\mathrm{\,cm^{-3}\,kpc}^{3\beta}$ where $\beta=0.54$ and assuming a Doppler width of $150 \mathrm{\,km\,s}^{-1}$, the optical depths towards Galactic center and anti-center at the line centroid ($18.96\,\mathrm{\AA}$) are $2.45$ and $0.45$, respectively, indicating that most photons are scattered more than once along the path. 
Moreover, if the hot gas is less turbulent, e.g., the entire contribution for the Doppler width is from the thermal motion of the plasma at $\approx50 \mathrm{\,km\,s}^{-1}$, the optical depth will be greater than $1$ at all directions. 
If the halo is truly turbulent, then a mean scattering term will not be valid. 
In either case, the optical depth effect should be included in any determination of the gas density distribution. 
The main purpose of this work is to incorporate a Monte Carlo radiative transfer (MCRT) simulation to reproduce the emission map and compare it with observations. 
In this way, we are able to constrain all model parameters to better precision. 
Moreover, we add into the hot gas model other components such as the rotation of the halo and a disk-like component, that have been discussed in the literature \citep{2016ApJ...822...21H,2009ApJ...690..143Y,2016ApJ...828L..12N}.

The paper is structured as follows. In Section 2, we describe all model assumptions including the density profile and rotation profile of the hot gas halo. In Section 3, we explain the data reduction and introduce the simulation method. We present the results from the simulation and the improved constraints on the density profile in Section 4 and discuss them in Section 5.

\section{Model}
The analyses of the hot gaseous halo rest on the model assumptions. Here we explain all premises we make based on previous observations, which include the temperature, density, rotation profile and metallicity model.
\subsection{Galactic X-ray Emission}
In this work, we are trying to study the hot gas component of the Milky Way by extracting its information from the soft X-ray emission observation. However, the data are comprised of several constituents beyond the hot gaseous halo. Two local contributors of the soft X-ray background (SXRB) are the solar wind charge exchange process (SWCX) and the Local Bubble (LB). SWCX is X-ray emission which occurs when highly ionized metals from the solar wind ($\mathrm{O}^{7+},\mathrm{O}^{8+}$ in our case) capture electrons from the neutral gas they interact with, either within the Earth's magnetosheath \citep{2003JGRA..108.8031R} or the heliosphere \citep{2004A&A...418..143L}. For the former mechanism, it is dependent on the solar wind activity and therefore temporal variations on timescales of hours to days, while the latter varies more on direction rather than time. 
The method utilized to deal with SWCX is discussed in \ref{sec: dataselection}. We have also known for decades the existence of a low-density cavity which is an old superbubble in which the Sun resides \citep{1987ARA&A..25..303C,1997ApJ...485..125S}. This Local Bubble (LB) is highly ionized at a  temperature $\sim 10^6\,\mathrm{K}$ \citep{1990ApJ...354..211S} which makes it a contributor to the SXRB. However, the structure of the LB is still unclear. 
Evidence supports either a volume-filled bubble \citep{1990ApJ...354..211S,2007PASJ...59S.141S} or a wall of gas \citep{2009Ap&SS.323....1W} at the edges of the bubble. In this work we adopt the constant-density volume-filled model used in MB15 and the details will be discussed in Section \ref{correction}. 

\subsection{Temperature}
Due to the sensitivity of the ion fraction of \ion{O}{7} and \ion{O}{8} in the $10^{6-7}\,\mathrm{K}$ temperature range \citep{1993ApJS...88..253S}, the temperature estimations of the Milky Way hot gas reach good agreement at about $\log T=6.2-6.3$ despite different measurements and model assumptions \citep{2004ApJ...617..232M,2005ApJ...624..751Y,2007ApJ...665..247W,2009ApJ...690..143Y,2010PASJ...62..723H,2012ApJ...756L...8G}. 
The best probe of the temperature structure comes from \cite{2013ApJ...773...92H}, which supports an isothermal temperature model. 
The authors use a subset ($110$ sight-lines) of emission lines in \cite{2012ApJS..202...14H} to minimize contaminations from SWCX, and process the data in a similar way to \cite{2012ApJS..202...14H} and fit the spectra with thermal plasma models for the hot gas halo. This work concludes a median temperature at $2.22 \times 10^6 \mathrm{K}$ with an interquartile range of $0.63 \times 10^6\mathrm{K}$ while the emission measure and intrinsic $0.5-2.0\,\mathrm{keV}$ flux varies by over an order of magnitude. Here we choose an intermediate value of $2\times10^6\,\mathrm{K}$.

Theoretical emissivities of \ion{O}{7} and \ion{O}{8} are obtained from AtomDB version $2.0.2$ \citep{2012ApJ...756..128F}, which assumes an APEC thermal plasma in collisional ionization equilibrium \citep[CIE, ][which is also assumed through out our model]{2001ApJ...556L..91S} and solar abundance \citep{1989GeCoA..53..197A}, which is $N_\mathrm{O}/N_\mathrm{H}=8.5\times10^{-4}$, whereas we adopt $N_\mathrm{O}/N_\mathrm{H}=5.5\times10^{-4}$ \citep{2001AIPC..598...23H}. Therefore we correct the emissivity obtained from AtomDB by a factor of $5.5/8.5=0.65$. At $2\times10^6\,\mathrm{K}$, the emissivities in units of $\mathrm{10^{-15} photons\,cm}^3\mathrm{\,s}^{-1}$ are $\epsilon_{\mathrm{O\,{V II}}}=3.09, 0.546, 2.395, 0.373$ for r, i, f, $\mathrm{He} \beta$ lines and  $\epsilon_{\mathrm{O~{V III}}}=1.45$. 

Since {\it XMM-Newton} EPIC-MOS is unable to resolve the \ion{O}{7} triplet as well as \ion{O}{8} resonance and \ion{O}{7} $\mathrm{He}\beta$, we simulate the photons of the triplet to fit the \ion{O}{7} data and the photons from the other two channels to fit the \ion{O}{8} data. Though there are no optical depth effects for the \ion{O}{7} forbidden line and intercombination line due to their small oscillator strengths.

\subsection{Density Profile}
The density profile of the hot gaseous halo can be derived from the Navarro-Frenk-White model assuming an isothermal temperature profile and be well approximated by the King profile  \citep[or $\beta$-model,][]{2010gfe..book.....M}
\begin{eqnarray}
n(r) = n_0 (1+(\frac{r}{r_c})^2)^{-3\beta/2}. \label{eq:beta-model}
\end{eqnarray}
Given its success in describing early-type galaxies \citep{1985ApJ...293..102F} and late-type galaxies\citep{2012ApJ...755..107D} as well as our Milky Way \citep{2015ApJ...800...14M,2016ApJ...828L..12N}, our simulation will mainly focus on this model. However, based on the absorption of X-ray binaries \citep{2005ApJ...624..751Y} and spectral analysis of diffuse gas along the LMC X-3 sight line \citep{2009ApJ...690..143Y}, an exponential disk structure with a height scale of about $1\sim3\mathrm{\,kpc}$ has been suggested. 
A model combining the halo and a thin disk ($<1\,\mathrm{kpc}$) is also favored by a joint study of low galactic and high galactic absorption lines \citep{2016ApJ...828L..12N}. 
We begin by exploring the spherical $\beta$-model on the all-sky emission and consider the disk structure as a modification to the gas density.

This model is applied in MB15 and they obtain a set of best-fit parameters $n_0r_c^{3\beta}=0.79\pm0.10\times10^{-2}\,\mathrm{cm}^{-3}\mathrm{kpc}^{3\beta}$, $\beta=0.45\pm0.03$ for \ion{O}{7} and $n_0r_c^{3\beta}=1.50\pm0.24\times10^{-2}\,\mathrm{cm}^{-3}\mathrm{kpc}^{3\beta}$, $\beta=0.54\pm0.03$ for \ion{O}{8}. They only consider a power-law approxomation $n(r)=n_0r_c^{3\beta}r^{-3\beta}$ without fitting the core radius because the lack of sight-lines towards the Galactic center. To compare with the previous work, we adopt the complete 3-parameter density model and determine each simultaneously.

\subsection{Rotation}\label{sec:rotation}
The hot halo is rotating in the same direction as the Galactic disk, base on observation of X-ray absorption lines against bright AGN continua, projected around the sky \citep[obtained contraint on the radial velocity $v_r=-15\pm20\,\mathrm{km\,s}^{-1}$ and the azimuthal velocity $v_{\phi}=183\pm41\,\mathrm{km\,s}^{-1}$]{2016ApJ...822...21H}.  In this work, the azimuthal velocity is fixed at $v_{\phi}=180\,\mathrm{km\,s}^{-1}$ beyond $1\mathrm{kpc}$ and linearly drops to $0$ within $1\mathrm{kpc}$ while the inflow velocity is neglected.

The rotation of the hot gas halo influences X-ray emission lines in two ways: 
(1) The rotation of the halo shifts the line centroid, but this effect is irrelevant in our simulation as the spectra are not resolved, and
(2) the line profiles are broadened and become non-Gaussian at lower latitude and directions away from the Galactic (anti-) center. 
 the broadening effect can be described by an effective Doppler-$b$ parameter, defined as $b=\sqrt{2\sigma^2}$. A rotationally broadened profile reduces the net optical depth along the line of sight and this effect can therefore be characterized in the radiative transfer simulation.

\begin{figure}
\epsscale{1.25}{}
\plotone{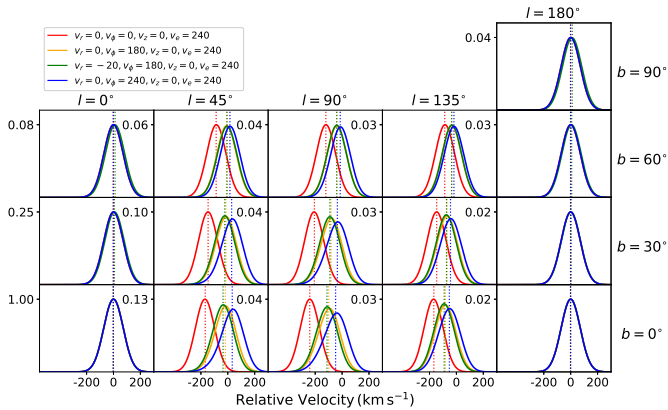}
\caption{The numerical prediction for X-ray emission spectra across the sky. The rotation models are same with those in \cite{2016ApJ...818..112M} and are color-coded in the same way described in the legend, where $v_r, v_\phi, v_z$ are the radial, azimuthal and axial velocity of the halo in units of $\mathrm{km\,s^{-1}}$ and $v_e$ is the speed of the Earth. The red curve stands for the stationary case. The additional orange curve is the model used in the simulation. The parameters are $\beta=0.5$, $r_c=2.5\mathrm{kpc}$, $b=100\mathrm{km\,s}^{-1}$. The vertical axes are rescaled so that the peak intensity is $1$ at $l=b=0^\circ$ \label{fig:emission_line_map}}
\end{figure}

\cite{2016ApJ...818..112M} calculated absorption lines for different rotation models of the hot Galactic halo. Here, we present in Figure \ref{fig:emission_line_map} the counterpart of those discussions for emission lines. In Figure \ref{fig:emission_line_map}, we assume the Doppler width $b=100\mathrm{\,km\,s}^{-1}$, and the result shows that the rotation effect is greatest roughly at midway between the Galactic center and the anti-center and decreases at higher latitudes.

\subsection{Other parameters}
Observations of the hot extended halos of external galaxies suggest a sub-solar abundance \citep[e.g.,][]{2016MNRAS.455..227A}. Cosmological simulations \citep{2002MNRAS.335..799T,2006ApJ...650..560C}, observations on external spirals \citep{2009ApJ...697...79R,2016MNRAS.455..227A} and high velocity clouds \citep{2000AJ....120.1830G,2004ASSL..312..195V,2005ApJ...630..332F} agree on a metallicity of $0.3\,\mathrm{Z_{\odot}}$. We adopt this value since it has successively been used to interpret data in \cite{2012ApJ...756L...8G,2013ApJ...770..118M,2015ApJ...800...14M,2016ApJ...828L..12N}.

We adopt a solar distance $d_{\odot}=8.5\,\mathrm{kpc}$ from the Galactic center \citep{2008ApJ...689.1044G}, along with the commonly accepted value $R_\mathrm{vir}=250\,\mathrm{kpc}$ and $M_\mathrm{vir}=1.5\times10^{12}\,M_\odot$ for the virial radius and the virial mass.

\section{Method} \label{sec:method}
\subsection{Data Reduction}
The data used here are almost identical to those in MB15 which are based on the catalog of {\it XMM-Newton} data produced by \cite{2010ApJS..187..388H,2012ApJS..202...14H}. Therefore, we only briefly summarize the data reduction process as these former works contain detailed descriptions. In the following two subsections, we review the data selection and emission line measure analysis conducted in \cite{2010ApJS..187..388H,2012ApJS..202...14H}. We also describe the additional filtering process MB15 use in Sec \ref{screen}. In Sec \ref{correction}, we describe the correction method applied to alleviate the effects from non-halo sources.

\subsubsection{Data selection}\label{sec: dataselection}
The original data were from the catalog of {\it XMM-Newton} prior to August 2010, containing 5698 observations with MOS exposures, and were processed with {\it XMM-Newton} Science Analysis System (SAS) version 11.0.1 (including the {\it XMM-Newton} Extended Source Analysis Software XMM-ESAS). Observing time affected by soft proton flaring were removed, which is distinguished by the count rate of the $2.5-12.0\,\mathrm{keV}$ light curve at that interval differing from the mean value (fitted by a Gaussian) by more than $1.5 \sigma$. Any observation with good observing time less than $5\,\mathrm{ks}$ or that does not have at least one MOS1 exposure and one MOS2 exposure was also discarded. Consequently, 2611 of the original 5698 observations remained after this screening procedure.

The authors conducted a source removal procedure by using the data from {\it XMM-Newton} Serendipitous Source Catalogue. Any point source within the field of view with flux in $0.5-2.0\,\mathrm{keV}$ band $F^{0.5-2.0}_X \geq 5 \times 10^{-14} \mathrm{\,erg\,cm}^{-2}\mathrm{\,s}^{-1}$ was excised by a circle of radius $50''$, which encloses $\approx 90\%$ of the source flux. In addition, bright sources not adequately removed by automatic processes and CCDs in anomalous states were excluded through visual inspection.

To reduce the contamination by SWCX, the authors used OMNIWeb, whose data are from satellites measuring in situ solar wind (mainly, {\it Advanced Composition Explorer (ACE)} and {\it Wind} ), and excluded the time when the SWCX is prominent (indicated by solar wind proton flux exceeding $2\times 10^{8} \, \mathrm{cm}^{-2}\,\mathrm{s}^{-1}$). This screening reduced the useful exposure times of some observations below the $5\,\mathrm{ks}$ threshold described above, which reduced the usable sight lines to $1435$.  
\subsubsection{Emission line measure}
Prior to extracting the emission line measurements, the authors subtracted the soft X-ray background (SXRB), including the SWCX, the Galactic and extra-galactic emission, the quiescent particle background (QPB) and the residual soft proton contamination, from the full field of view. The measurements were conducted with XSPEC version 12.7.0. 
For each observation, the authors fitted the data in $0.4-10.0\,\mathrm{keV}$ band with a multicomponent spectral model including two delta functions for the \ion{O}{7} and \ion{O}{8} $K_{\alpha}$ emission with the centroid for \ion{O}{7} left as a free parameter; the energy of \ion{O}{8} is fixed at $0.6536\,\mathrm{keV}$ \citep[from APEC;][]{2001ApJ...556L..91S}. 
This method measured all oxygen line emission including that from the (residual) SWCX, LB emission and attenuated halo emission. 
The Galactic and extra-galactic emission were modeled with an absorbed APEC thermal plasma model and an absorbed power-law (EPL) model with a photon index of $1.46$, respectively. The authors also used a power-law to model the residual contamination from the soft proton, though part of this effect was reduced through the cleaning method introduced above. 
The APEC and EPL component were attenuated using the XSPEC absorption model \citep{1992ApJ...400..699B,1998ApJ...496.1044Y} with \ion{H}{1} column data from the LAB \ion{H}{1} survey \citep{2005A&A...440..775K}. 

\cite{2012ApJS..202...14H} measured the statistical error and considered the uncertainties contributed from APEC and EPL fitting and combined the two in quadrature. To quantitatively account for the soft proton contamination, the authors calculated the total flux in $2-5\,\mathrm{keV}$, $F_{\mathrm{total}}^{2-5}$, and its EPL component, $F_{\mathrm{exgal}}^{2-5}$. The ratio of these two values was treated as a threshold and any observation with the ratio below $2.7$ was discarded. The final screening ruled out the soft-proton-contaminated observations and reduced the number of usable data from 1435 to $1003$.

\subsubsection{Further Screening}\label{screen}
MB15 applied further screening methods to obtain the most reliable set of observations. They removed observations close to possible X-ray sources including bright X-ray sources ($ROSAT$-BSC), galaxies (PGC 2003), galaxy clusters (MCXC) and quasars ($ROSAT$-RLQ;-RQQ).

Some Galactic X-ray structures are also contaminations that should be avoided when characterizing the extended hot halo, such as supernovae and superbubbles. The large Galactic absorption correction at low latitudes can be rather uncertain, due to the absence of an all-sky H$_2$ map. Therefore, MB15 excluded the observations with Galactic latitude $l \leq 10^{\circ}$. 
The Fermi Bubble \citep[FB,][]{2010ApJ...724.1044S} at the center of the Galaxy also exhibits strong X-ray emission, thus the observations with $|l|\leq22^{\circ}$ and $|b|\leq55^{\circ}$ were excluded. 
The region of the Fermi Bubbles is treated separately in \cite{2016ApJ...829....9M}. Finally, they removed a cluster of data near the Large Magellanic Cloud and the Small Magellanic Cloud.

After all filtering processes, MB15 had $649$ sight-lines and conducted analyses to obtain best fits for the halo density profile. However, when they compared their best fit model with \ion{O}{8} data, they found one observation ({\it XMM-Newton} ObsID $0200730201$, $(l, b) = 327.59^{\circ}, +68.92^{\circ}$) was $9\sigma$ above the model prediction ($I_{\mathrm{obs}} = 8.69\,\mathrm{L.U.}$ compared to $I_{\mathrm{mod}}=1.18\,\mathrm{L.U.}$, $\mathrm{L.U.}$ is the line units: $\mathrm{photons}\,\mathrm{cm}^{-2}\,\mathrm{s}^{-1}$) and the goodness-of-fit can be significantly improved from $\chi^2_{645}=1.21$ to $\chi^2_{644}=1.08$ if this observation is excluded. This excess of emission might due to the supernovae located within the north polar spur and should not be considered in our hot gas halo model.
Though MB15 only exclude it for the \ion{O}{8} analysis and used all $649$ sight-lines for \ion{O}{7}, we use only $648$ observations for both \ion{O}{7} and \ion{O}{8} lines.
\subsubsection{Residual Corrections}\label{correction}
The major Galactic components that must be corrected for when studying the hot halo include the emission from the FB and LB, the extinction effect of the disk and the SWCX process. The effect of FB has been eliminated by avoiding the sight-lines through the bubbles and the effect of the disk is partially alleviated in the same way. Regarding the Galactic absorption, considerable extinction can occur on individual sight-lines, though sight-lines from the disk (i.e., $|b|<10^{\circ}$) are excluded. The local bubble with a temperature slightly below that of the halo contributes to the emission line as well. Consequently, the observed intensities can be decomposed as
\begin{eqnarray}
I_{\mathrm{obs}} =I_{\mathrm{halo}}e^{-\tau(N_{\mathrm{H~I}})} + I_{\mathrm{LB}}.
\end{eqnarray}  
The \ion{H}{1} column and extinction rate at different lines of sight are obtained from HEASARC \footnote{\url https://heasarc.gsfc.nasa.gov}, whose data are based on \ion{H}{1} surveys \citep{1990ARA&A..28..215D,2005A&A...440..775K}. Considerations for the LB are not as straight-forward and the contribution in the optically thin case is
\begin{equation}
	I_{\mathrm{LB}} =\frac{\epsilon_{\mathrm{LB}}}{4\pi}\int_0^L n^2\mathrm{d}\,s.
\end{equation}
However, the geometry of the local bubble is still unclear.  
We adopt the model in MB15, where they use a constant-density, volume-filled local bubble model with a uniform temperature and the boundary of the local bubble in every direction varies in the range $100-300\,\mathrm{pc}$ \citep{2003A&A...411..447L}. Therefore, the emissions from the local bubble become a linear function of its path length,
\begin{eqnarray}
I_{\mathrm{LB}}=\frac{\epsilon_{\mathrm{LB}}}{4\pi}n^2L.
\end{eqnarray}
The path length of each sight-line is inferred from $1003$ \ion{Na}{1} absorption line equivalent width measurements in \cite{2003A&A...411..447L}. 
MB15 leave the constant of the local bubble density as a free parameter in the fitting process. Here, we use their best fit values $n=4.0 \times 10^{-3}\,\mathrm{cm}^{-3}$ for \ion{O}{7} and $n=0.7 \times 10^{-3}\,\mathrm{cm}^{-3}$ for \ion{O}{8}. We also use the same temperature assumption that the LB is at $\log T(K) = 6.1$  which is lower than the halo temperature by a factor of $2$. 

\subsection{Modeling Optical Depth Effects}\label{sec: opticaldepth}
MB15 fit a power-law density model to the emission data, but they do not determine the core radius independently from the normalized density parameter due to the lack of observations near the Galactic center. Their correction for the optical depth effect is limited to a simplification of the radiative transfer function -- a single-scattering model. However, as is shown later, multi-scattering processes contributes a non-negligible part and affects the surface brightness distribution across the sky. In this section, we explain the principles of the radiative transfer code and how parameter choices affect our comparison with data.

\subsubsection{Monte Carlo Radiative Transfer}\label{ch:principle}
The essence of the simulation is to reproduce the radiative transfer process in the halo. 
The simulation focuses mainly on the propagation rather than the mechanisms of emission and absorption, therefore, those physical processes are treated statistically in the context of photon. 
The emission is through collisional excitation of the CIE gas. 
Photons are emitted randomly in angle and subject to a radial distribution proportional to the squared density profile of the halo,
\begin{eqnarray}
p(r) = C r^2 n^2(r),
\end{eqnarray}
where $C$ is a normalization. This distribution can be sampled, for any set of parameters, through an acceptance-rejection Monte-Carlo method \citep[][]{flury1990acceptance}.

The history of a photon is determined by the radiative transfer process, which is given by 
\begin{equation}
	\frac{\mathrm{d}\,I_{\nu}}{\mathrm{d}\,s} = -\kappa_{\nu} I_{\nu},
\end{equation}
where $I_{\nu}$ is the specific intensity at a given frequency and $\kappa_{\nu}$ the extinction coefficient. The emission and scattering term does not appear because we are tracking every single photon rather than doing a field analysis.

In terms of discrete photons, the specific intensity can be written as \citep{2011BASI...39..101W}
\begin{equation}
I_{\nu} = \frac{h\nu N_{\nu,\mathbf{\Omega}}}{\mu\Delta\mu\Delta\phi\Delta\nu\Delta A},\label{intensity_defination}
\end{equation}
where $h\nu$ is the energy of the photon; $\mu = \cos\theta$ is the cosine of the latitudinal angle and the solid angle $\mathrm{d}\,\Omega=\mu\mathrm{d}\,\mu\,\,\mathrm{d}\,\phi$; $A$ is the cross area. Consequently, the solution to the above equation is, in terms of photon number in a certain parameter range, 
\begin{eqnarray}
N_{\nu,\mathbf{\Omega}}(\tau_{\nu,\mathbf{\Omega}}) = N_0e^{-\tau_{\nu,\mathbf{\Omega}}}, \label{eq:2}
\end{eqnarray}
where $\mathrm{d}\,\tau_{\nu,\mathbf{\Omega}}=\kappa_{\nu}\mathrm{d}\,s(\mathbf{\Omega})$ is the optical depth. This solution can be understood in two equivalent ways. First, a group of photons run synchronously but the survival number of photons after each step (spatial or temporal) obeys an exponential distribution dictated by the optical depth. Alternatively, a photon is stopped at a certain distance (absorbed or scattered), but this path length, in terms of optical depth, varies from one photon to another and is subject to the exponential distribution. In simulations, the latter interpretation is more computationally efficient and is adopted in this work. 

At the beginning of the streaming of every photon, a random number subjected to the exponential distribution with an expectation of unity is sampled, which is interpreted as the optical depth of the free-path of this photon. To determine the termination of the photon, the optical length is converted into geometric distance and fulfilled step by step. The smaller step the photon takes, the better accuracy of the terminal position will be achieved. Here we choose a typical step at $0.3\,\mathrm{kpc}$ and adaptively scaled by a factor $(r/r_c)^{3\beta}$ if the photon is far from the center where the plasma is very diffuse.  The conversion from optical depth to geometric length is calculated through 
\begin{eqnarray}
\mathrm{d} \, \tau_{\nu} = \frac{\pi e^2}{m_e c}n_{X} f\phi(\nu)\,\mathrm{d}\,s, \label{eq:dtau2ds}
\end{eqnarray}
where the first term is the constant $0.02654\,\mathrm{cm}^2\,\mathrm{s}^{-1}$, $n_X$ the number density of absorbers and $f$ the oscillator strength. $\phi(\nu)$ is the normalized line shape related to the Doppler parameter $b$ by
\begin{eqnarray}
\phi(\nu) = \frac{1}{\sqrt{\pi}\nu_D}e^{-\big(\frac{\nu-\nu_0}{\nu_D}\big)^2} \,\,\,\,\,\,\,\nu_D=\frac{b}{c}\nu_0, \label{eq:lineprofile}
\end{eqnarray}
where $\nu_0$ is the laboratory frequency and $c$ the speed of light.

For resonance lines, the photons are not destroyed but are absorbed and re-emitted.  
The re-emission is isotropic relative to the absorber and at a different frequency, based on the local Gaussian line shape. 
For a $0.5\mathrm{\,keV}$ X-ray photon, the Compton scattering cross section is insignificant relative to the absorption cross section. 
Therefore there is no practical difference between scattering and absorption.

The simulation begins with a photon created somewhere inside the halo according to the density profile Eq.\ref{eq:beta-model}. The photon propagates repeatedly according to Eq. \ref{eq:2} unless it is detected by the ``telescope'' or escapes the boundary of the halo. The total number of photons created and the number of photons detected gives the intensity as a function of angle, which is compared to the data.
When collecting the photons, the geometric symmetry can be exploited to promote efficiency. For the stationary model (non-rotating gaseous halo), the system is spherically symmetric. Therefore every photon that reaches the sphere $8.5\,\mathrm{kpc}$ from the Galactic center (distance of the Sun from the Galactic center) is indistinguishable from each other and can be collected directly. The simulation effectively obtains the flux by counting the number of the photons reaching the Solar Sphere, and the flux should be divided by the projection term to obtain the specific intensity ($\mu=\cos\theta$ in equation \ref{intensity_defination}.
 In this case where the detection surface is a sphere, the projection angle $\theta$ is the angle between the sight-line and the Galactic center).

In the case of a rotating halo, the cylindrical symmetry makes the simulation much more computation intensive. The ``telescope'' is a ring that lies on the Solar circle. Consequently, the projection onto the plane introduces a factor $\cos\theta=\sin b$ where $b$ is the Galactic latitude.

\subsection{Data Fitting}
We compare the data of the 648 sight-lines with our MCRT simulation. For a given set of simulation parameters ($\Theta$) the fitness is described by $\chi^2$ defined as
\begin{eqnarray}
\chi^2(\Theta) = \sum_{\mathrm{i}=1}^{648} \frac{(I_\mathrm{i}-s_\mathrm{i}(\Theta))^2}{\sigma_i^2},
\end{eqnarray} 
where $I_\mathrm{i}$ and $\sigma_\mathrm{i}$ are the corrected intensity and error of the data; $s_\mathrm{i}$ is the intensity given by the simulation at the same location. The connotation of ``same location'' is different for stationary model and rotating model.

For the stationary model, we contract the coordinates of the observation into its angle towards the Galactic center $\theta$ and multiply them by the factor $\cos\theta$ as the conversion from specific intensities to fluxes, and $2\pi\sin\theta$ as the integral over the degenerate azimuthal angle. The collected photons are binned by $100$ equal-width-bins along the $\theta$-axis, and smoothed by a polynomial function to estimate their $1\mathrm{D}$ distribution. The $s_\mathrm{i}$ values are chosen at the same $\theta$ location of the smoothed histogram.

For the co-rotating model, we make $2\mathrm{D}$ bins for both the $648$ sight-lines and the simulated emission map. Due to the mirror symmetry with respect to the $b=0$ plane, $s_\mathrm{i}$ is calculated by the weighted mean density of the photons within the bin of the i$_{\mathrm{th}}$ data sight-line and the 8 bins around it and the mirror images of these $9$ bins.

The parameters of interest are the three profile parameters, $n_0$, $\beta$, $r_c$, and the Doppler factor $b$. For the oxygen ions at $T=2.0 \times 10^6 \mathrm{K}$, the thermal motion contributes $b_\mathrm{th}=50\,\mathrm{km\,s}^{-1}$ to the Doppler width. Here we are only interested in the non-thermal part of the Doppler width, therefore we sample $b_\mathrm{turb}$ and convert it into $b_{\mathrm{tot}}=\sqrt{b_\mathrm{turb}^2+b_\mathrm{th}^2}$ in the simulation. To find the best combination of these parameters that is most consistent with data, we adopt the Monte-Carlo Markov chain (MCMC) method by using the Python package {\tt emcee} \citep{2013PASP..125..306F} to explore the parameter space. At each MCMC step, a set of parameters is chosen and the corresponding simulated data are generated through the  MCRT and the log-likelihood $-0.5\chi^2$ is calculated to evaluate the goodness of the set of parameters. 

\section{Results} \label{sec:result}
In this section we summarize the results from the MCRT-MCMC fitting. Similar to the method in MB15, we fit the two sets of ion lines separately  because the \ion{O}{7} lines have a higher mean S/N (4.9) while that of \ion{O}{8} is lower (1.3).

\subsection{Stationary Model Parameter Estimation}
\begin{figure}[h!]
\epsscale{2}
\plottwo{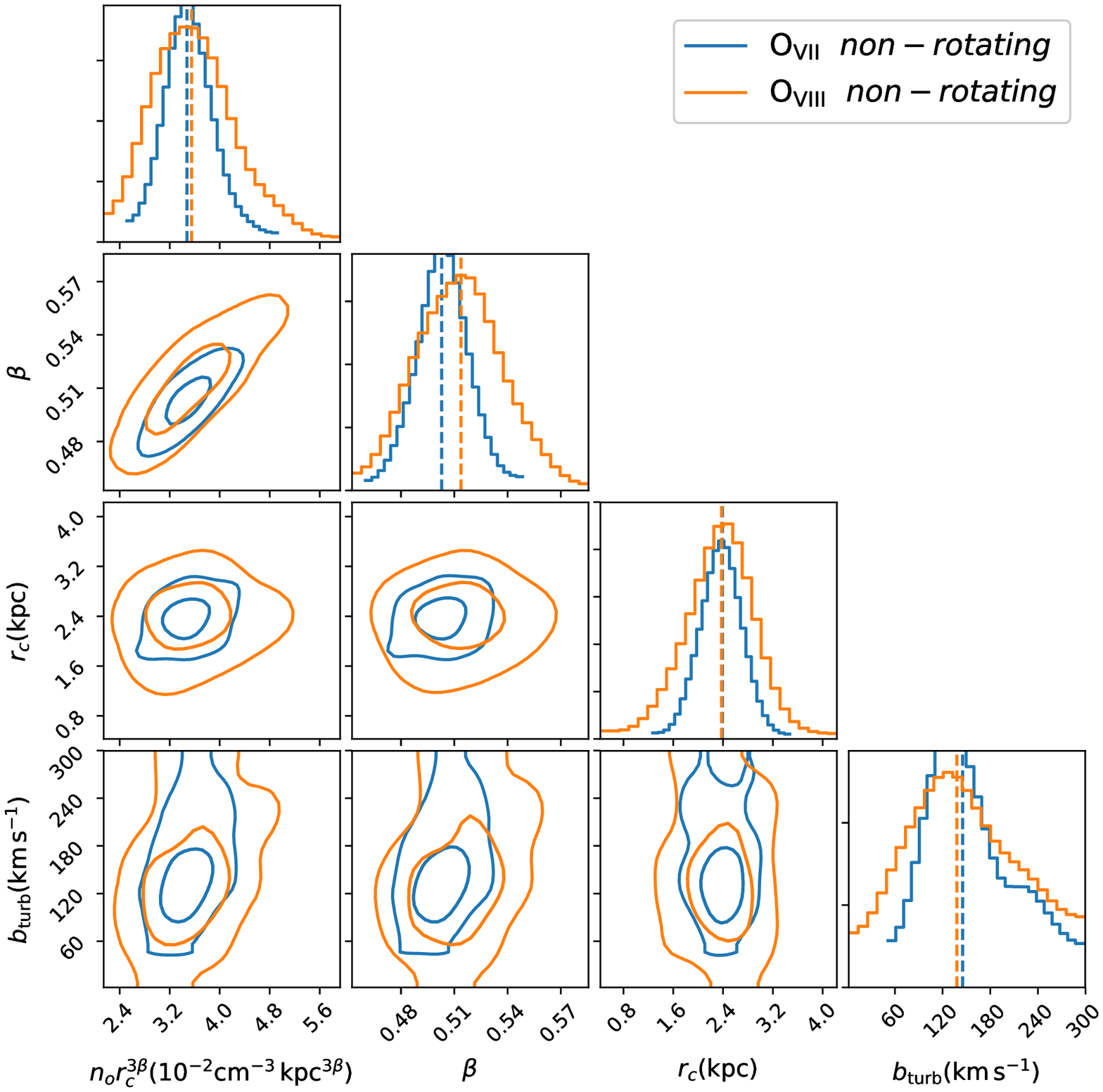}{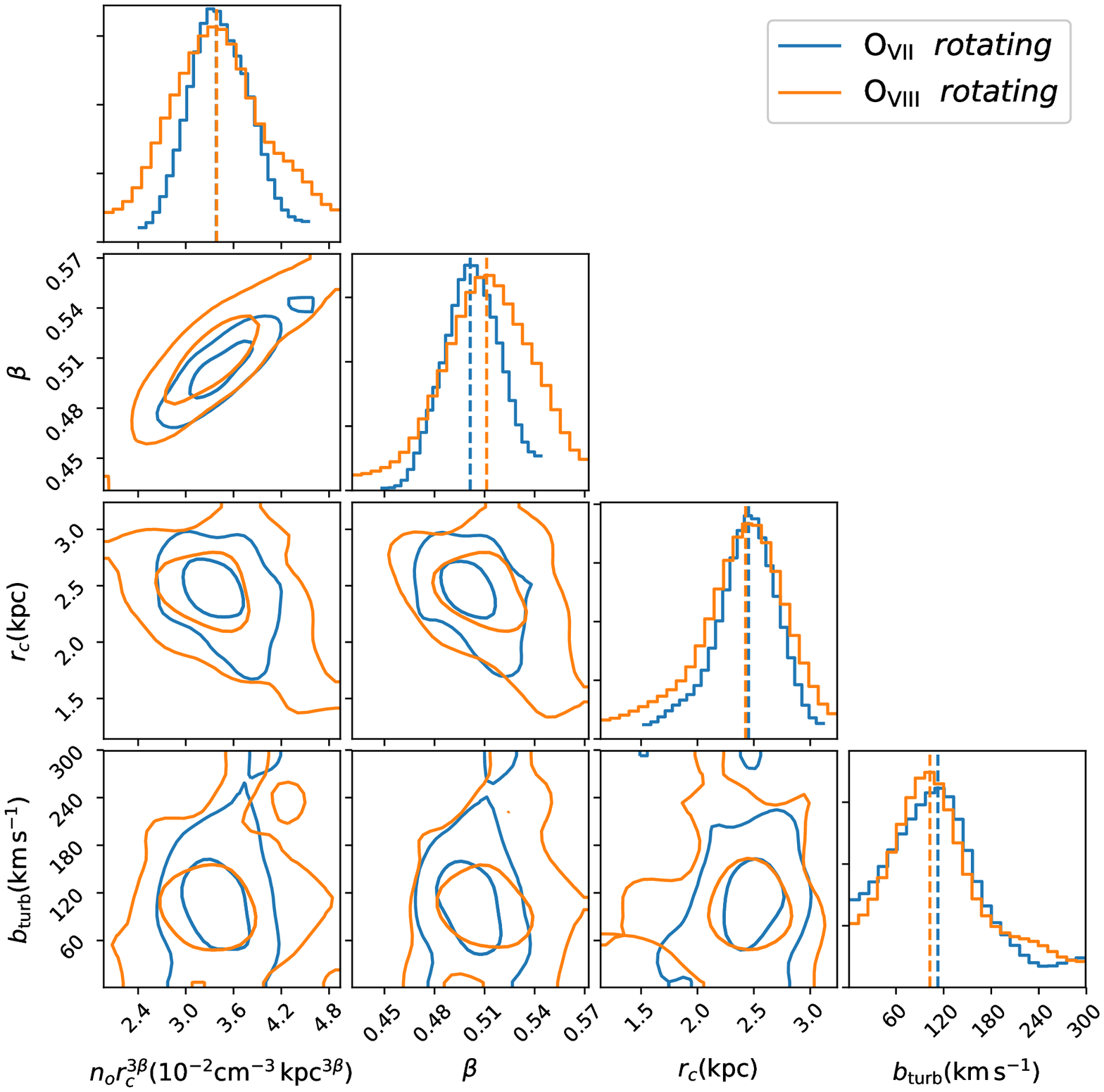}
\caption{MCMC model parameters log-likelihood surface for \ion{O}{7}(blue) and \ion{O}{8}(orange) in the stationary model (upper) and rotating model(lower) with $v_\phi = 180 \,\mathrm{km\,s^{-1}}$. The contours are for the $1\sigma$ and $2\sigma$ ranges which correspond to $39\%$ and $86\%$ confidence level in the 2D case.  \label{fig:mcmccontour}}
\end{figure}

For the stationary model, we use $10^5$ photons in each MCMC step to obtain a simulated emission map with high S/N, and in turn, a relatively robust parameter estimation. The results for stationary model fitting are plotted in the upper panel of Figure \ref{fig:mcmccontour} and listed in Table \ref{tab:1} (No. 1,4). In order to compare the results with those in MB15, we use $n_0r_c^{3\beta}$ (instead of $n_0$) as the normalization, even though we do not use a simplified power-law to approximate equation \ref{eq:beta-model}. As a result, we see a correlation in the best fit between the normalization and $\beta$ in Figure \ref{fig:mcmccontour}. 
This is due to the degeneracy of the two parameters
since $\beta$ and $n_0$ have opposite effects on the density at a fixed radius. 
The flatness (i.e., $\beta$) of the halo is crucial in determining the total baryon content of the halo. For the stationary case, we have $\beta$ constrained at $0.50\pm0.02$, which is consistent with, if not slightly higher than, the results from emission line studies \citep[][]{2015ApJ...800...14M,2016ApJ...822...21H} and lower than those from absorption line studies assuming similar profiles \citep[][and model A of \citealt{2016ApJ...828L..12N}]{2013ApJ...770..118M}.
The core radius is also constrained at $r_c=2.4^{+0.3}_{-0.3}$ for \ion{O}{7} and $r_c=2.4^{+0.4}_{-0.5}$ for \ion{O}{8} which are consistent with the value $2.1\sim2.5\,\mathrm{kpc}$ found by \cite{2016ApJ...828L..12N}. This result comes as a surprise because the sight-line in our data closest to the Galactic center has an angle of about $30^{\circ}$, while a core radius at $\sim 2.4\,\mathrm{kpc}$ corresponds to an angle of about $16^{\circ}$. The fourth plot on the left panel of Figure \ref{fig:mcmccontour} shows the log-likelihood surface of the non-thermal part of the Doppler parameter, $b_\mathrm{turb}$; both \ion{O}{7} and \ion{O}{8} results indicate that the plasma is turbulent at a speed of about $>140\,\mathrm{km\,s}^{-1}$ (near the speed of sound at $200\,\mathrm{km\,s}^{-1}$). There is also a weak degeneracy between $n_0r_c^{3\beta} $ and $b_\mathrm{turb}$ in the model because the two parameters act oppositely in determining the optical depth (Eq: \ref{eq:dtau2ds} and \ref{eq:lineprofile}). 

Even though the parameter estimations for the both ions are close, there is a  difference in the goodness of the fitting, which was also a problem in MB15. In the stationary model, we have 4 parameters to be explored (all shown in Figure \ref{fig:mcmccontour}), therefore the degree of freedom in the $\chi^2$ fitting is 644. Hereafter, we use the reduced $\chi^2_{644}$ ($\chi^2$ per degree-of-freedom) as the indicator of the goodness-of-fit. For \ion{O}{8}, the best-fit gives a $\chi^2_{644} = 1.16$ , which is acceptable, whereas the $\chi^2_{644}$ for \ion{O}{7} is as high as $5.06$. The way that MB15 use to lower the $\chi^2$ to an acceptable value is to add an additional variation, $\sigma_{\mathrm{add}}=2.1 \mathrm{\,L.U.}$ to the \ion{O}{7} lines; a similar technique is also used in \cite{2013ApJ...770..118M} for their \ion{O}{7} absorption line study. The deviation can be caused by variations in the SWCX, in the density/temperature profile, in the LB emission, or in the optical depths. According to the analyses in their paper, the SWCX postulation can be ruled out since they did not find an emission excess near the ecliptic plane and there was little difference between the fitting using only the sight-lines near the ecliptic plane and the sight-lines in the opposite directions. The variation of the halo density/temperature profile and the LB emission are both plausible since the models are quite idealistic in both cases. We do not attempt to distinguish these scenarios and use the additional $2.1\,\mathrm{L.U.}$ variation. The $\sigma_{\mathrm{add}}$ is added to the statistical and systematic uncertainties \citep[from ][]{2012ApJS..202...14H} in quadrature.   

\subsection{Co-rotating Model Parameter Estimation}

The efficiency of the simulation drops by about an order of magnitude when rotation is introduced. Therefore we use fewer photons ($10^4$) in each MCMC step. Nevertheless, the MCMC procedure still gives reliable parameter estimation. 
The presence of the rotation affects the radiative transfer in the halo and in turn modifies the parameter estimation; we present the contours on the lower panel in Figure \ref{fig:mcmccontour}. The differences between the normalization and $\beta$ remain while the estimation for the core radius reaches better agreement at $\approx2.4\mathrm{kpc}$. The primary difference between the stationary case and the co-rotating case is the non-thermal Doppler $b_\mathrm{turb}$ estimation which decreases to $\approx 110\mathrm{km\,s}^{-1}$ for both ions in the presence of rotation. As shown in Table \ref{tab:1}, the estimations of the $b_\mathrm{turb}$ for co-rotating model (No.2,6,7) are systematically lower than that of the stationary model (No,1,4,5). Combining this with the results from Figure \ref{fig:emission_line_map}, we see that this effect can be explained by the line broadening effect of the halo rotation.

However, the rotation model is not necessarily a better description of the observation in terms of $\chi^2_{644}$ which is $4.97$ and $1.16$ for \ion{O}{7} and \ion{O}{8} respectively. Therefore, we also add the variation $\sigma_{\mathrm{add}}=2.1\,\mathrm{L.U.}$ to \ion{O}{7} data, making sure that the parameter estimation is not biased by a few sight-lines with small intrinsic uncertainties. 

\section{Discussion} \label{subsec:conclusion}
\subsection{Consistency with the absorption line data}
\begin{figure}[!h]
\epsscale{2.5}
\plottwo{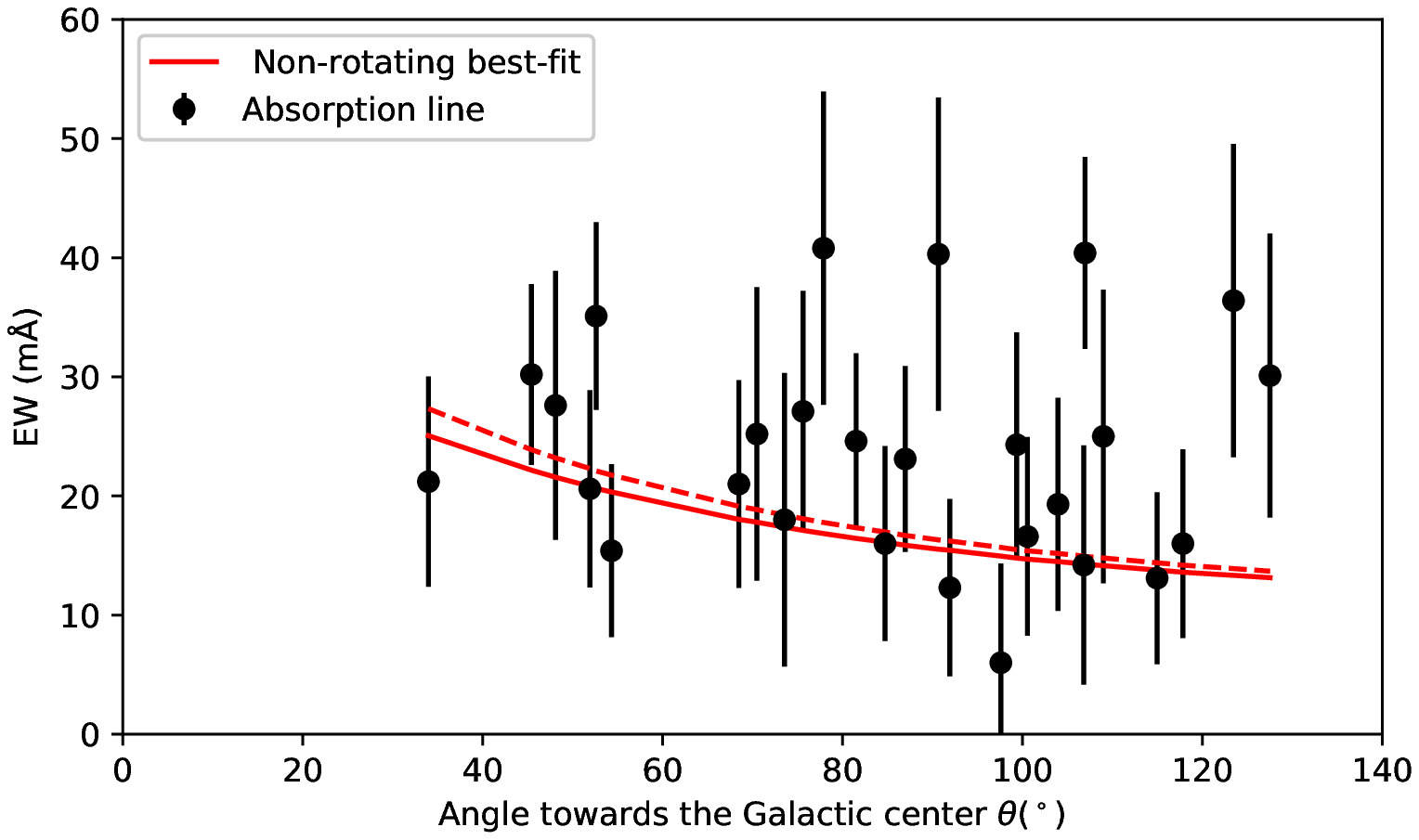}{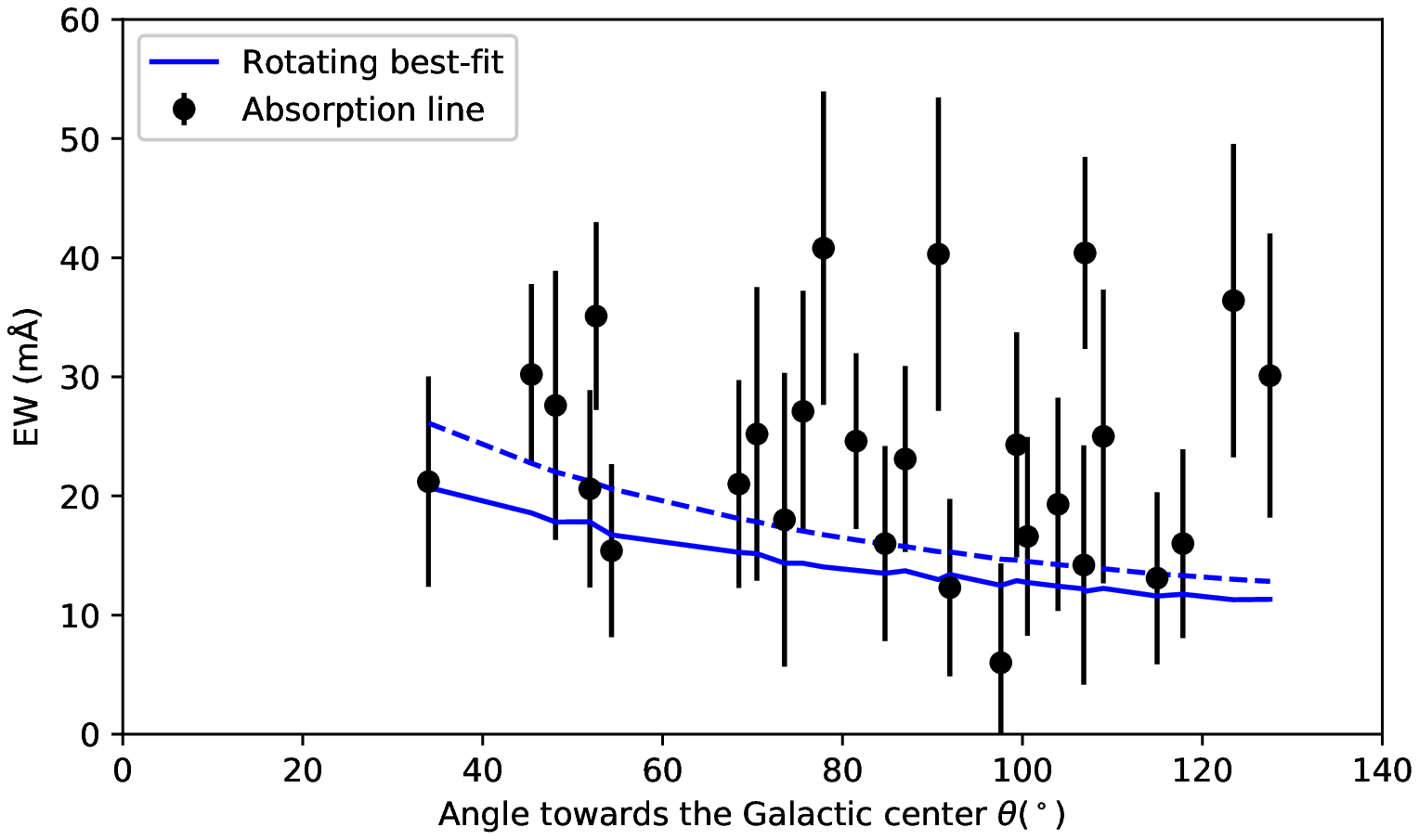}
\caption{The equivalent width (EW) comparison between the data and the model. The data is a subset of the data (27 of 37 sight-lines) in Table 1 of \cite{2016ApJ...822...21H} with 10 sight-lines with highest uncertainties removed for illustration purpose. The EW measurement has very small uncertainties therefore we likewise adopted an additional uncertainty of $7.5\,\mathrm{m\AA}$ in the EW of each sight-line as it was introduced in \cite{2013ApJ...770..118M} to account for the ignorance of the intrinsic variance due to the substructure of the absorbing medium. The red and blue solid lines show the EW towards same sight-lines for model No.5 and 7. (also with additional uncertainties of $2.1\mathrm{\,L.U.}$), with the dashed lines being their $2\sigma$ upper limit.\label{fig:EW}}
\end{figure}

The different mechanisms of emission and absorption along with their different dependencies on the density of the ions make the cross check between the observation on emission lines and absorption lines very important in revealing the structure of the hot gaseous halo \citep[e.g.,][]{2012ApJ...756L...8G}. We calculate the equivalent width (EW) of our best-fit models in directions where the absorption line data are available \citep{2016ApJ...822...21H} and present the result in Figure \ref{fig:EW}. It appears that the model built upon the emission line data underestimates the absorption line EW values by about $36\%$ (as measured by the median value). However, this result is not inconsistent with the EW measurement in the way that the EW data with larger values generally have larger uncertainties. Therefore we find it not very helpful in improving the $\chi^2$ calculation regard to the EW data by tuning the density normalization of the model (i.e., shifting the model curves upward to meet the majority of the EW data).

\subsection{Optical Depth Effects}
\begin{figure}[!h]
\epsscale{1.15}
\plottwo{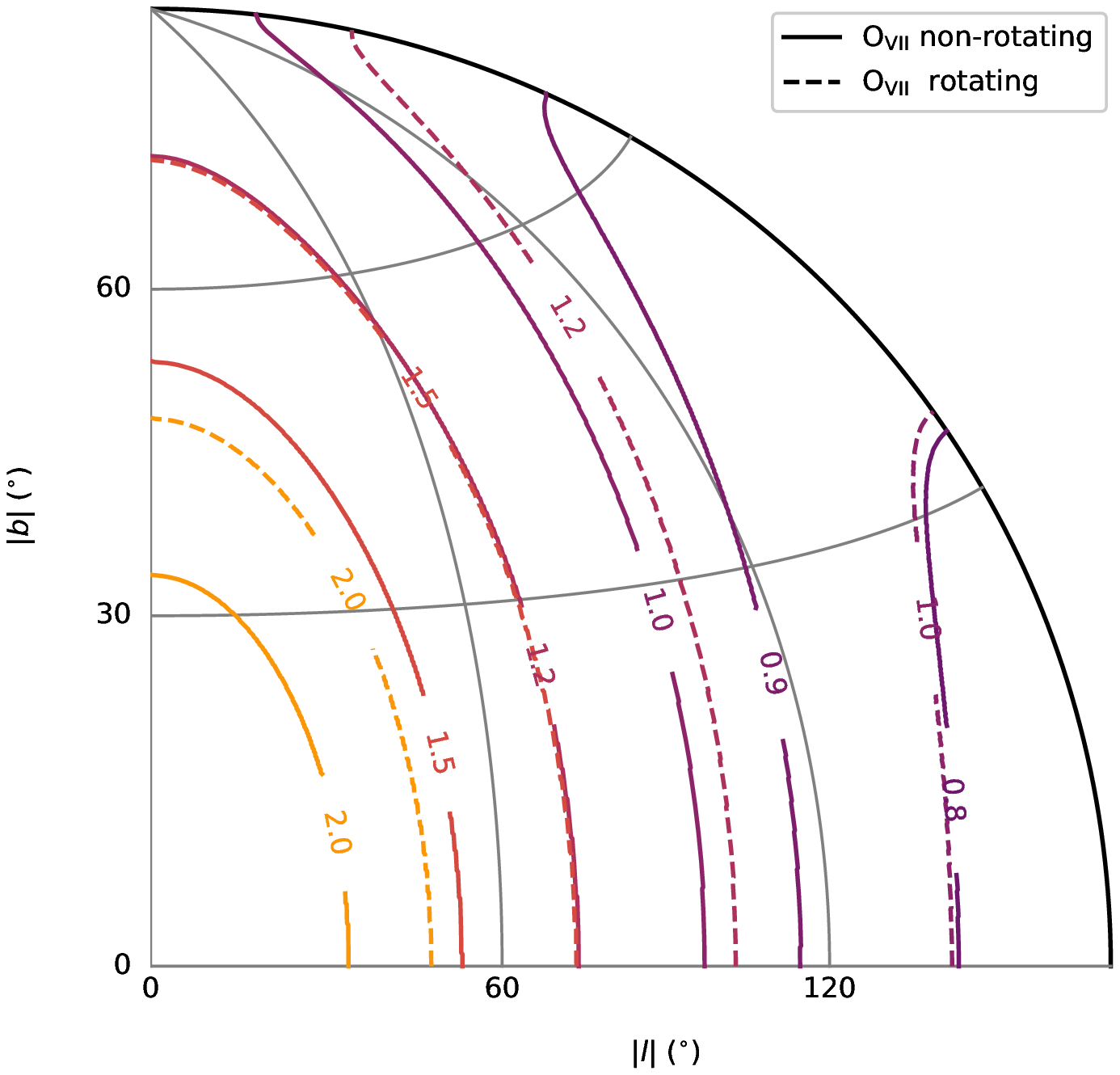}{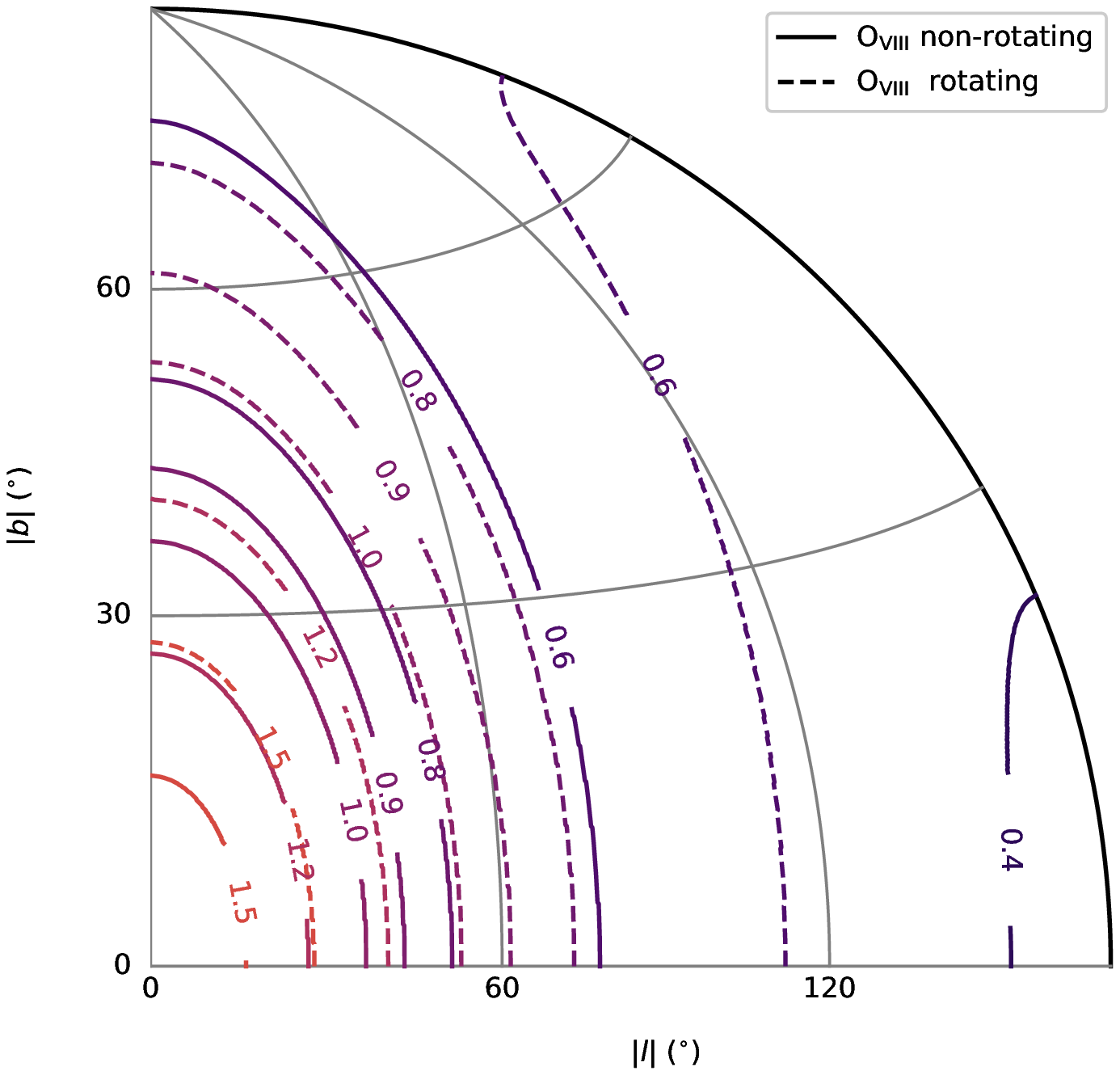}
\caption{The \ion{O}{7} and \ion{O}{8} line center optical depth contour across the sky. We only consider the resonance line for the two ions at $21.6\,\mathrm{\AA}, 18.9\,\mathrm{\AA}$ for their higher optical depths over intercombination lines and forbidden lines (\ref{sec: opticaldepth}). The parameters are obtained from the best-fit of the stationary model and the rotating model (No.1,2,4,6).}\label{fig: tau-contour}
\end{figure}

We are able to map the optical depth for the two kinds of lines across the sky with the best-fit parameters. The optical depth contours for \ion{O}{8}  and \ion{O}{7} emission lines are presented in Figure \ref{fig: tau-contour}. For \ion{O}{8} lines, the optically thick area approximately overlaps with the Fermi Bubble while the \ion{O}{7} optical depth is greater than unity towards all directions except the Galactic anti-center. The rotation model (dashed lines) gives slightly higher values of the optical depths than the stationary case (solid lines), mainly because the rotation reduces the $b_\mathrm{turb}$ by attributing it to the broadening effect induced by the rotation.\\ \par

\begin{figure}
\epsscale{1.3}
\plotone{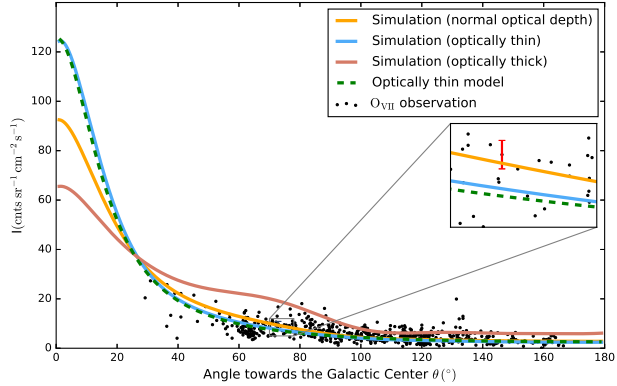}
\caption{The surface brightness distribution as the function of the angle towards the Galactic center $\theta$ for the stationary halo model. The orange, blue and brown solid lines are from simulation with different Doppler parameter at $110\,\mathrm{km\,s}^{-1}$, $1000\,\mathrm{km\,s}^{-1}$ and $5\,\mathrm{km\,s}^{-1}$, representing a normal optical depth case and cases at optically thin and thick limits ($\tau=1.2, 0.13, 26$ at $b=90^\circ$ respectively). To compare the simulation with semi-analytical result, we directly calculate the surface brightness at different directions, using the emission measure formula for optically thin case: $I\propto\int n^2 \mathrm{d}\,s$, and plotted in green dashed-line. The \ion{O}{7} data are also included as black dots for reference and the typical length of the error bar is shown in red in the zoom-in plot. Other parameters of the halo profile are $n_0r_c^{3\beta}=3.0\times 10^{-2}\mathrm{cm}^{-3}\mathrm{kpc}^{3\beta}$, $\beta=0.5$, $r_c=2.5\,\mathrm{kpc}$. \label{fig:1d_surface_brightness}}
\end{figure}

The surface-brightness as the function of angle from the Galactic center is presented in \ref{fig:1d_surface_brightness} for different optical depth models.
In optically thin limit, the simulation (blue) and analytic model (green) match well, while for a normal optical depth, the (orange) curve is damped near the Galactic center and raised at directions far from the center. 
As the scattering is characterized by the density of the ions, which is higher at the vicinity of the Galactic center, more photons from the Galactic center are scattered out of the sight-line and re-distributed to larger detection angle. 
The difference between the models can be distinguished by the data, leading to a good determination of the optical depth effects. 

We also examined a case where the optical depth is approximately $20$ times larger than our best-fit case. This is achieved by artificially reducing the Doppler b parameter to $5 \mathrm{\,km\,s^{-1}}$, which is about an order of magnitude lower than the thermal value at the virial temperature.
The simulated surface-brightness profile is plotted in brown, which shows a flattening and subsequent decrease at about $60\arcdeg-90^\circ$. In this case, the free-path length of the photon is less than  $1\mathrm{\,kpc}$ towards the Galactic anti-center ($\tau \ga 15$) that the local density profile have a significant contribution to the emission distribution. As the optical depth approaches much larger values, the mean free path becomes much shorter and the emission becomes isotropic.

The optical depth effects redistribute the emission intensities across the sky and affect the parameter estimation of the halo model. By conducting Monte-Carlo radiative transfer simulations, we find that the general effect is re-distributing the intensities at lower angles (either Galactic longitudes/latitudes or angles towards the Galactic center) to larger angles such that the curve of the surface-brightness is flattened when optical depth effects are considered. However, this flattening profile is similar to an optically thin plasma with a smaller value of $\beta$ (a flattened density profile). Leading to some degeneracy, as seen in the elongated confidence contours between $n_0r_c^{3\beta}$ and $\beta$ in Figure \ref{fig:mcmccontour}.

\begin{figure}[!h]
\plotone{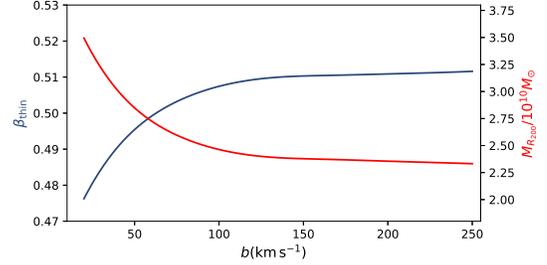}
\caption{Results of fitting simulated emission maps with different Doppler $b$ parameters with optically thin models. The best-fit $\beta_{\mathrm{thin}}$ is plotted in blue while the enclosed baryon mass of the hot halo within the virial radius is plotted in red.}\label{fig: beta-b}
\end{figure}

We simulate the \ion{O}{7} emission at the $648$ sight-lines to quantitatively investigate how optical depth effects modulate the $\beta$ estimation. 
The simulated data are fit with optically thin models assuming the total Doppler $b$ parameter ranging from $20-250\,\mathrm{km\,s}^{-1}$ (only to show the effects of different $b$, though $b<50\,\mathrm{km\,s}^{-1}$ is less than the thermal width and therefore not realistic for the hot halo). 
The fitting is carried out though a MCMC routine, leaving the $\beta_{\mathrm{thin}}$ and the normalization to be free parameters and the core radius is fixed at $2.5\,\mathrm{kpc}$. 
The results of the $\beta_{\mathrm{thin}}$ are plotted as blue curve in Figure \ref{fig: beta-b}. At lower Doppler $b$, where the optical depth effects are prominent, the fit leads to an underestimation of $\beta$. The underestimation problem becomes less significant with increasing $b$ values, which is closer to an optically thin plasma. In turn, the underestimation of the slope of the halo causes an overestimation in the baryon budget estimation (red curve in Figure \ref{fig: beta-b}). 

\subsection{Turbulence}
We set constraints on the turbulence $b_\mathrm{turb}$ of the gaseous halo by carrying out radiative transfer simulation. For the stationary model fitting, we obtain systematically higher $b_\mathrm{turb}$ for both ion lines ($\sim 150\,\mathrm{km\,s^{-1}}$). However, when the galactic rotation is included in the model, we see that the value drop to $90\sim120\,\mathrm{km\,s^{-1}}$. This can be explained by the line-broadening effect of the rotation as mentioned in Sec \ref{sec:rotation}. 
The total Doppler-$b$ in the co-rotating case would be $100\sim130\,\mathrm{km\,s^{-1}}$ which is consistent with the results in \cite{2012ApJ...756L...8G,2015ApJS..217...21F,2016ApJ...828L..12N}. However, our result is a global model which includes the co-rotation of the halo, while the previous ones are based on sight-line to sight-line observations. It is also interesting to point out that, \cite{2016ApJ...828L..12N} analyze the low galactic latitude and high galactic latitude absorption lines and find that the former have a larger average Doppler-$b$ $125\,\mathrm{km\,s^{-1}}$ compared to the latter $95\,\mathrm{km\,s^{-1}}$. 
This can be explained by our rotation-induced-broadening effect \citep[][ and Figure \ref{fig:emission_line_map}]{2016ApJ...829....9M} and can be further examined by high spectral resolution line profile analyses or a longitudinal study.

If we regard the $b_\mathrm{turb}$ estimation obtained from the rotating-halo model as the true turbulence of the hot gaseous halo, we have $b_\mathrm{turb}\approx100\mathrm{km\,s}^{-1}$. With the speed of sound of the halo $c_\mathrm{s}=220\mathrm{km\,s}^{-1}$ we obtained the pressure contribution from the two components,
\begin{eqnarray}
\frac{P_\mathrm{th}}{P_\mathrm{turb}} = \frac{\rho c_\mathrm{s}^2/\gamma}{\rho\sigma^2} = 1.2\times\frac{c_\mathrm{s}^2}{b^2_\mathrm{turb}} \approx 5.8,
\end{eqnarray}
where $\gamma=\frac{5}{3}$ is the adiabatic index. This value is smaller than the result from the circumgalactic medium simulation \citep[about $12$,][]{2017MNRAS.466.3810F}, but it is consistent with the general picture of massive galaxy formation \citep{2003MNRAS.345..349B} that the halo can maintain the hydrostatic equilibrium with the thermal support. This shows a particular level of feedback, which can be important to modelers.

\subsection{Disk-like Component}
The spherical halo model is justified by the fitting, but some observations of absorption lines towards individual sources 
have been fit with a disk-like halo \citep{2005ApJ...624..751Y,2009ApJ...690..143Y,2010PASJ...62..723H} with a scale height of a few $\mathrm{kpc}$ or less \citep{2016ApJ...828L..12N}. Additionally, observations of low ionization lines are also successfully interpreted with an exponential disk model. This disk component is a natural consequence of supernova events in the MW disk. The outflow driven by the supernova feedback can maintain a disk-like halo at temperature $\sim 10^6\,\mathrm{K}$, but the detailed shape, extent and temperature of this halo strongly depend on the supernova history, temperature of the mid-plane  and radiative cooling.

\begin{figure}[!h]
\epsscale{2.0}
\plottwo{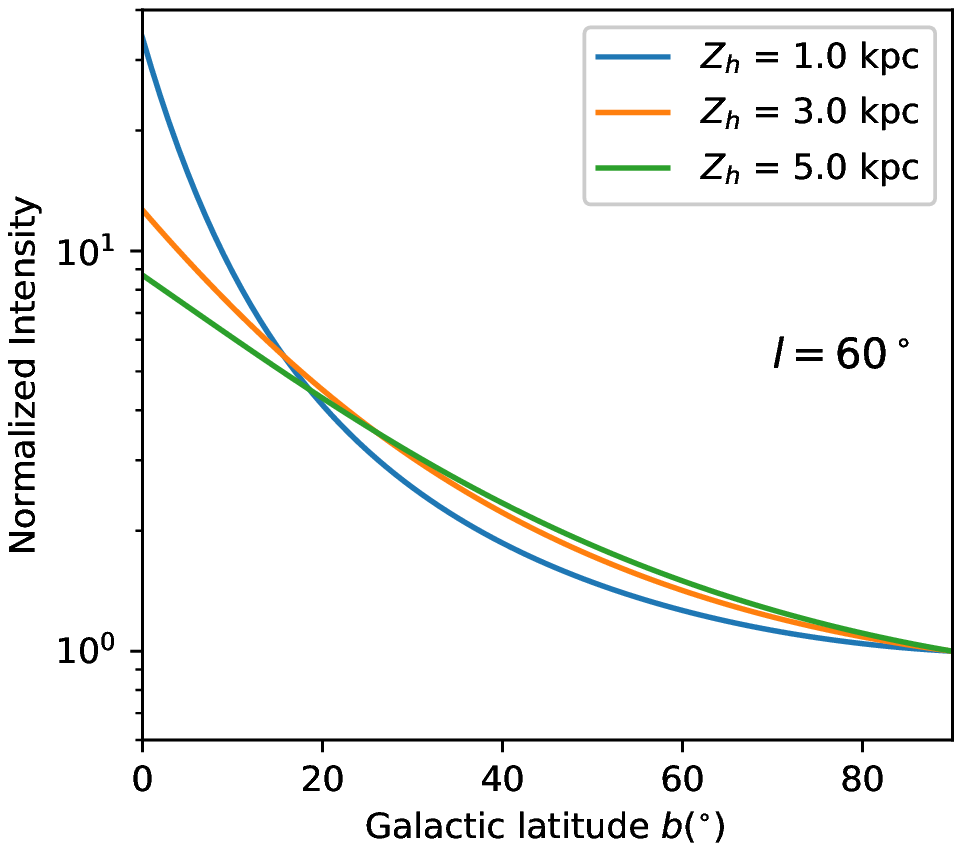}{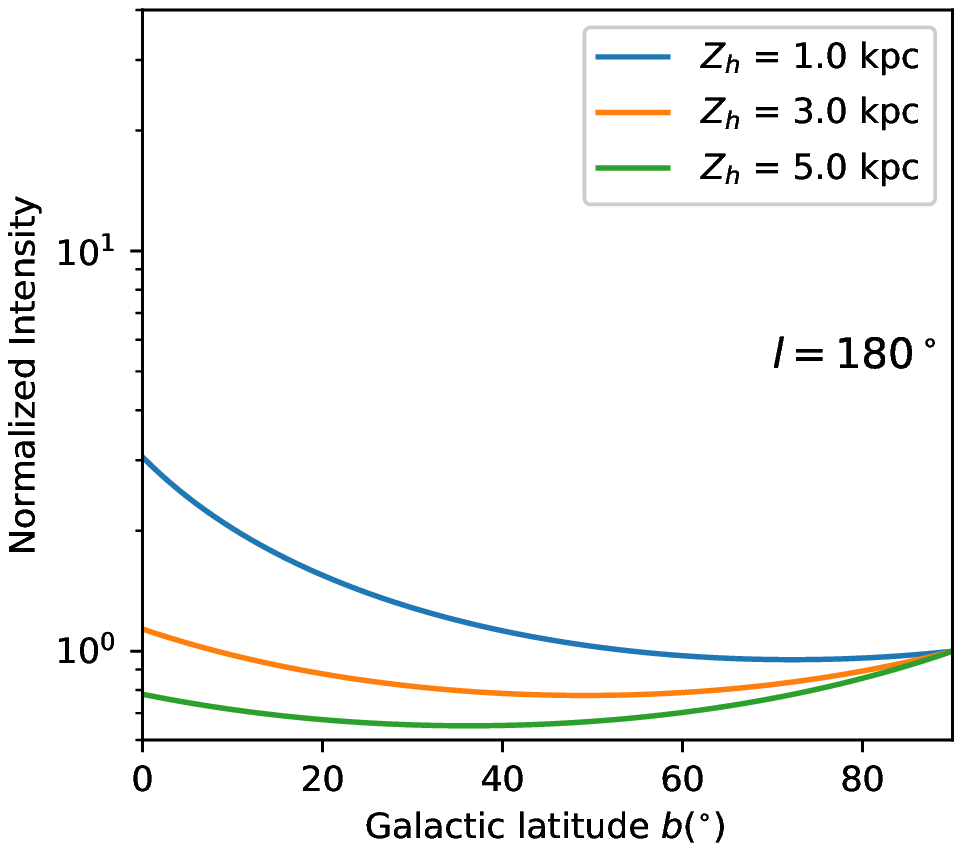}
\caption{The comparison between disk emission intensities at Galactic longitude $l=60^\circ$ and $l=180^\circ$ as a function of Galactic latitude for different $z_h$.  The curves are normalized at $b=90^{\circ}$ to illustrate the differences in latitudinal dependence.}
\label{fig:disk_feature}
\end{figure}

\cite{2016ApJ...822...21H} use the same MCMC method as MB15, but add an exponential disk in the fitting. Likewise, we add an exponential disk to both the stationary model and the co-rotating model, but we assume the density of the disk decreases exponentially both along the $z$ direction and the $r$ direction since we should not expect the disk can be sustained at larger radii where the MW disk fades out. 
The profile is
\begin{eqnarray}
n_{\mathrm{disk}}=n_{o,\mathrm{disk}} e^{-r/r_h}e^{-z/z_h}\label{eq: diskprofile}.
\end{eqnarray} 
This radial dependency of the disk density leads to distinct differences from the former disk model at certain directions. While the intensities towards $l=60^\circ$ decrease with Galactic latitudes, as we would expect from a normal disk model, the emission intensities can rise at high Galactic latitudes at $l=180^\circ$ for thick disks (Figure \ref{fig:disk_feature}). This makes the direction towards $l=180^\circ$ a useful probe for different disk models.

We first make a test with a pure disk model and find that it can not be fitted with \ion{O}{8} data. For \ion{O}{7} lines the $\chi^2_{646}=6.05$ is no better than the spherical halo models. Therefore, the disk component, if exists, should come as a complement of the spherical halo. For simplicity, we assume that the disk and the halo share the same metallicicy and temperature. We fix the scale radius at $r_h=3\,\mathrm{kpc}$ and leave $z_h$ to be a free parameter. The fitting is done only for the co-rotating model because of its cylindrical symmetry. The results are listed in table \ref{tab:1} (No. 3,8,9).

We show in Figure \ref{fig:alonglb} the emission measure of the \ion{O}{7} data at different longitudes and latitudes. The spherical halo model (red) underestimates the emission at lower latitudes, while a spherical halo with a disk component (blue) makes a better fit. In particular, the spherical halo model and the disk model can be distinguished by their performance along the latitudes at $l=180^\circ$ (the last panel in Figure \ref{fig:alonglb}) since we would expect from the spherical halo a higher emission measure at higher latitudes towards which the path length is greater, while the disk model ($z_h \sim 1$ $\,\mathrm{kpc}$) makes an opposite prediction. The \ion{O}{7} data in the last panel shows a decreasing pattern with increasing latitude and the spherical halo + disk model reduces the differences between the single halo model and the data (by a factor $\la2$ at lower $b$). The fitting is done globally and the improvement in one single panel is limited. The averaging shown in Figure \ref{fig:alonglb} enables one to better comprehend how different components influence the fit.

\begin{figure}[!h]
\epsscale{1.4}
\plotone{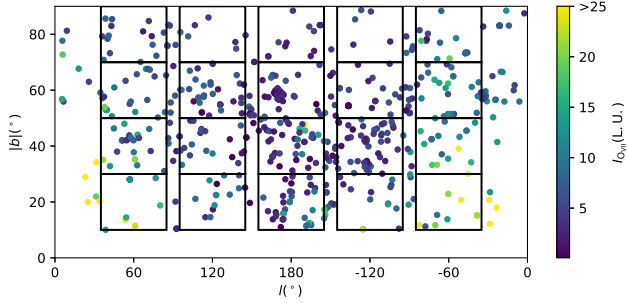}
\caption{The emission map of \ion{O}{7} and the bins used for statistics. The equal-size-bins are chosen to cover most of the sight-lines. The emission measure of each bin is assigned as the weighted mean of the intensity of the sight-lines inside the bin. The weight is defined as the inverse of the uncertainty.}
\label{fig:binmap}
\end{figure}

\begin{figure}[!h]
\epsscale{1.4}
\plotone{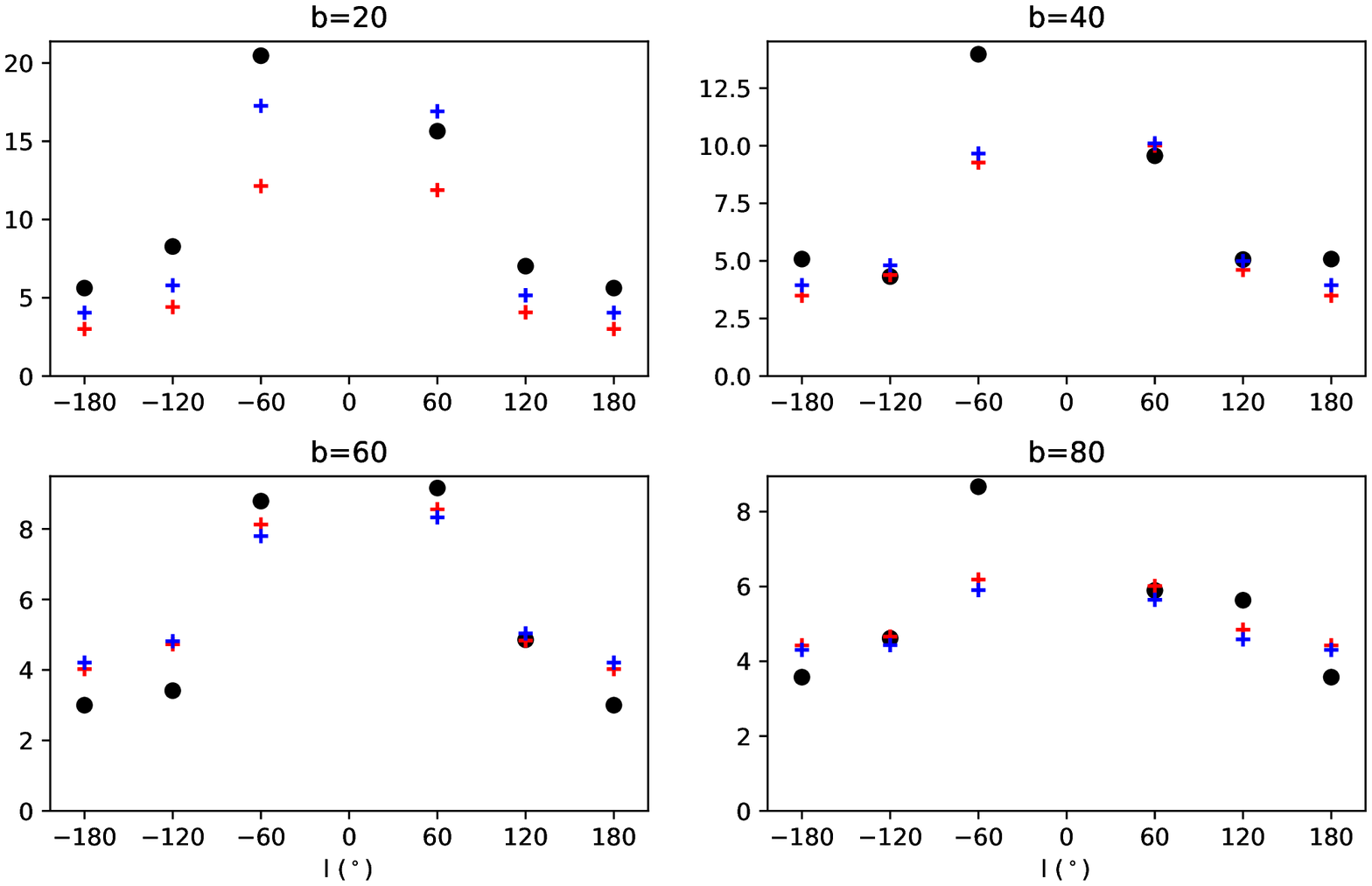}
\plotone{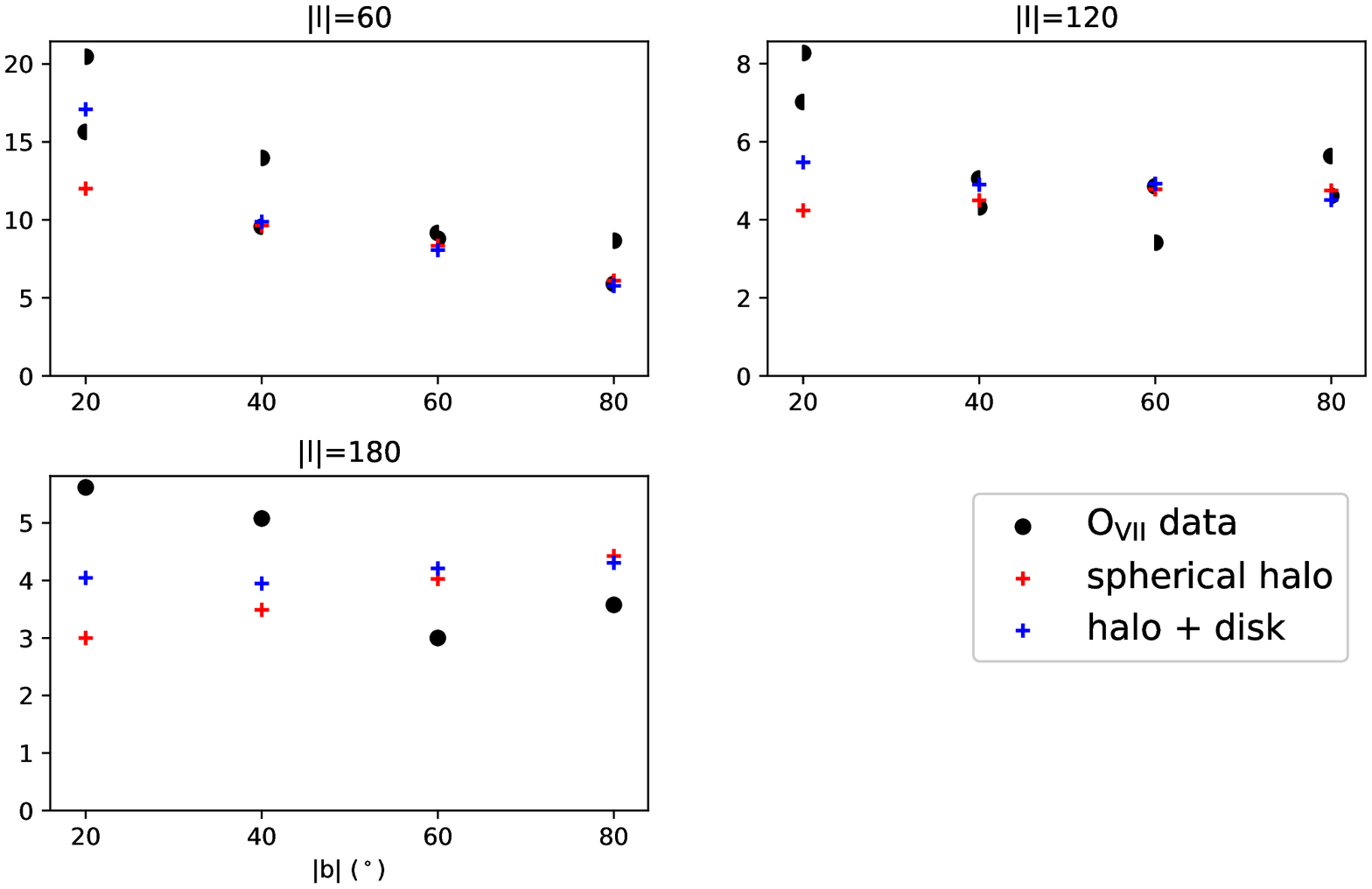}
\caption{The \ion{O}{7} emission intensities from the observations (black circles) and simulated emission map along the Galactic longitude ($l$) at different Galactic latitudes ($b$) (first 4 panels), and the emission along the Galactic latitude at different Galactic longitudes (last 3 panels). In the last 3 panels the intensities of the data towards positive and negative longitudes are denoted with left-filled and right-filled circles. The alignment of the bins are shown in Figure \ref{fig:binmap}. We simulated the emission map of two models: the spherical halo with rotation in red and the spherical halo with rotation and a disk in blue. The parameters of these two models are acquired through the fitting No.7 and No.9. The spherical halo with a disk fits the observation better in the way that it raises the emission measure at lower latitudes while the spherical halo model is unable to reproduce the latitudinal variation.}
\label{fig:alonglb}
\end{figure}

 The results of the fitting (Table \ref{tab:1}) show that the observation and model are brought closer in the presence of an exponential disk, and the scale height of the disk inferred from \ion{O}{7} lines is constrained within the range $\sim1.3 \,\mathrm{kpc}$ while that from \ion{O}{8} is less constrained.
 This disk is thinner than the results in \cite{2009ApJ...690..143Y,2010PASJ...62..723H}, because the disk in our model only contributes to a small portion of the emission. 
 Our disk is thicker than the $0.16\,\mathrm{kpc}$ disk from the low Galactic latitude line analyses in \cite{2016ApJ...828L..12N}, but the results are consistent in the sense that both disks contribute to a similar small portion of baryon mass $1.4\times10^8\,M_\odot$ (Table \ref{tab:1}). 
 It is worth pointing out that the spherical halo plus disk model in \cite{2016ApJ...828L..12N} finds a smaller $\beta=0.33$ than $\beta=0.62$ in the spherical $\beta$-model, while the extra disk component in our model improves the fit at lower Galactic latitudes without significantly affecting the $\beta$-model parameters. Therefore, though the disk mass in both models are small, they result in different total mass.

 The baryon mass contribution from the disk is 2 orders of magnitude lower that of the halo, which leads us to the conclusion that the disk only makes up a small portion of the ion column and this is also consistent with the previous z-exponential model \citep[$\approx 10\%$,][]{2016ApJ...822...21H}.
 This halo part of the result is also consistent with that of the halo+disk fitting for same emission line data \citep{2016ApJ...822...21H}, though we update their disk model and their scale height, which was not well constrained. 

\subsection{Baryon Budget}
\begin{figure}[!h]
\epsscale{1.3}
\plotone{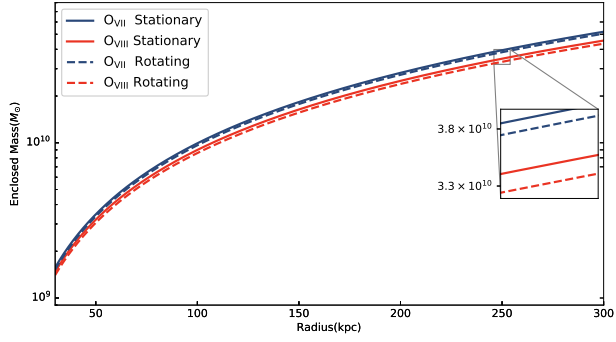}
\caption{The enclosed baryon mass of the halo versus radii inferred from the best-fit of different models. The metallicity used is $0.3Z_{\odot}$. The uncertainties can be found in Table \ref{tab:1}.}
\label{fig:mass}
\end{figure}

With the better constraint of the halo parameters, we update the baryon mass estimation of the Milky Way halo (Figure \ref{fig:mass}). We find that the gas mass estimations are not sensitive to the uncertainties in the parameters.  At the  virial radius of $250\mathrm{\,kpc}$, the enclosed mass is about $3.1^{+0.5}_{-0.3}\times 10^{10}\,M_{\odot}$ and can only account for $18\%$ of the missing baryon mass ($1.7 \times 10^{11}M_{\odot}$, after the exclusion of the contribution from the stars and cold gas.).
The data from the Galactic center and the FB are not used in modeling, and the density profile we obtain here might not properly describe those structures. Nevertheless, it is shown that the central region of the Galaxy is either depleted of gas \citep{2015ApJ...807...77K,2016ApJ...828L..12N} or contains a fractional mass compared to the halo mass \citep[$10^{7-8}\,M_\odot$,][]{2016ApJ...829....9M,2016ApJ...828L..12N}.
\par

Further improvements in interpreting the emission can be achieved with higher spectral resolution data, which would permit one to isolate lines and obtain better S/N for each line.  Such improved resolution is likely to occur from future missions with calorimeters, which have sufficient resolution to split the \ion{O}{7} triplet, of which two of the lines have low optical depth. Measuring each of the \ion{O}{7} triplet line has the potential of separating cooler and hotter gas, as well as more accurately separating the contribution from solar wind charge exchange and the Local Bubble.  Higher spectral resolution observations will allow one to determine line center shifts, which further constrain halo rotation and turbulent broadening. Higher S/N data will allow for better measures of the \ion{O}{8} Ly$\alpha$ line as well as recombination from higher level recombination lines, which have lower optical depths.  Together, these will lead to a more detailed and precise description of the Milky Way hot halo.  A calorimeter would be one of the main instruments on a replacement mission for Hitomi \citep{2016SPIE.9905E..0UT}, the Athena mission \citep{2016SPIE.9905E..2FB}, or the Lynx mission \citep{2015SPIE.9601E..0JG}.

\acknowledgments
We thank all the people that have offered help to this work, including Matthew Miller, Edmund Hodges-Kluck, Zhijie Qu and our anonymous referee whose comments helped to improve this work. Also special thanks to Eric Peng for the computational resources he provided. This research is funded by the China Scholarship Council, NASA ADAP grant NNX16AF23G, and the Department of Astronomy at the University of Michigan, who we also thank for their hospitality.

\software{AtomDB \citep{2012ApJ...756..128F}, {\tt emcee} \citep{2013PASP..125..306F}}
%




\begin{rotatetable}
\begin{deluxetable*}{ccccccccccccc}

\tablenum{1}
\tablecaption{Parameter estimation results}
\tablewidth{0pt}
\tablehead{
\colhead{} & No. &\colhead{$V_{\phi}(r>1\mathrm{kpc})$} &\colhead{$n_0r_c^{3\beta}$}&\colhead{$\beta$}&\colhead{$r_c$}&\colhead{$b_\mathrm{turb}$}&\colhead{$\sigma_{\mathrm{add}}$}&\colhead{$n_{\mathrm{0,disk}}$}&\colhead{$z_h$}&\colhead{$\chi_\nu^2(\mathrm{d.o.f.})$}&\colhead{$M_\mathrm{halo}(r\leq250\,\mathrm{kpc})$} & \colhead{$M_\mathrm{disk}(r\leq50\mathrm{\,kpc})$\footnote{The density is converted from the $I_\mathrm{0,disk}$ by assuming same temperature and metallicity of the disk with that of the halo.}}\\
& \colhead{} & \colhead{$\mathrm{km\,s}^{-1}$} &\colhead{$10^{-2}\mathrm{cm}^{-3}\mathrm{kpc}^{3\beta}$}&\colhead{}&\colhead{$\mathrm{kpc}$}&\colhead{$\mathrm{km\,s}^{-1}$}&\colhead{$\mathrm{L.U.}$}&\colhead{$10^{-2}\mathrm{cm}^{-3}$}&\colhead{$\mathrm{kpc}$} & &\colhead{$10^{10}M_{\odot}$} & \colhead{$10^8M_{\odot}$}\\
}\label{tab:1}

\startdata
\ion{O}{8} 
			&1& - & $3.55^{+0.65}_{-0.56}$ & $0.51^{+0.02}_{-0.02}$ & $2.38^{+0.40}_{-0.47}$ & $138^{+90}_{-65}$ & - & - & - & $ 1.16(644)$ & $ 3.5^{+0.4}_{-0.6}$ & -\\
			&2& $180$ & $3.39^{+0.67}_{-0.55}$ & $0.51^{+0.03}_{-0.03}$ & $2.43^{+0.30}_{-0.39}$ & $103^{+78}_{-46}$ & - & - & - & $1.16(644)$ & $3.3^{+0.6}_{-0.7} $ & -\\
            &3& $180$ & $2.86^{+0.62}_{-0.41}$ & $0.51^{+0.03}_{-0.02}$ & $2.44^{+0.48}_{-0.35}$ & $92^{+62}_{-46}$ & - & $3.62^{+2.73}_{-1.39}$ & $0.97^{+1.46}_{-0.68}$ & $1.12(642)$ & $2.8^{+0.3}_{-0.8} $ & $1.3^{+0.6}_{-0.7}$\\
\hline
\ion{O}{7} 
		   &4& - & $3.48^{+0.38}_{-0.31}$ & $0.50^{+0.01}_{-0.01}$ & $2.39^{+0.28}_{-0.31}$ & $145^{+76}_{-39}$ & - & - & - & $5.06 (644)$ & $3.9^{+0.2}_{-0.5} $ & -\\
           &5& - &  $3.48^{+0.36}_{-0.42}$& $0.50^{+0.01}_{-0.02}$ & $2.09^{+0.39}_{-0.55}$ & $224^{+56}_{-65}$ & $2.1$ & - & - &  $1.77 (644)$ &$3.9^{+0.5}_{-0.4} $ & -\\
           &6& $180$ & $3.38^{+0.42}_{-0.29} $& $0.50^{+0.02}_{-0.01}$ & $2.45^{+0.23}_{-0.28}$ & $113^{+50}_{-62}$ & - & - & - &  $4.97 (644)$ &$3.8^{+0.4}_{-0.6} $ & -\\
           &7& $180$ & $3.60^{+0.43}_{-0.33} $& $0.51^{+0.01}_{-0.01}$ & $2.30^{+0.34}_{-0.42}$ & $125^{+55}_{-64}$ & $2.1$ & - & - &  $1.74 (644)$ &$3.5^{+0.5}_{-0.3} $ & -\\
           &8& $180$ & $2.99^{+0.26}_{-0.32}$& $0.52^{+0.01}_{-0.01}$ & $2.67^{+0.28}_{-0.20}$ & $115^{+34}_{-24}$ & - & $3.99^{+1.31}_{-0.83}$ & $1.25^{+0.49}_{-0.44}$ &  $4.54 (642)$ &$2.5^{+0.5}_{-0.3} $ & $1.8^{+0.3}_{-0.4} $\\
           &9& $180$ & $2.82^{+0.34}_{-0.29}$& $0.51^{+0.02}_{-0.01}$ & $2.53^{+0.17}_{-0.18}$ & $110^{+47}_{-43}$ & $2.1$ & $3.82^{+1.03}_{-0.74}$ & $1.34^{+0.51}_{-0.43}$ &  $1.56 (642)$ &$2.8^{+0.5}_{-0.4} $ & $1.8^{+0.2}_{-0.3}$\\
\enddata
\end{deluxetable*}
\end{rotatetable}




\bibliography{citation}



\end{document}